\documentstyle[amstex,aps,psfig,epsfig]{revtex}

\begin{document}
\title{Precise Measurement of Dimuon Production Cross-Sections in $\nu _{\mu }$Fe
and $\bar{\nu}_{\mu }$Fe Deep Inelastic Scattering at the Tevatron}
\author{M.~Goncharov, T.~Adams, A.~Alton, T.~Bolton, J.~Goldman}
\address{Kansas State University, Manhattan, KS, USA}
\author{P.~Spentzouris, J.~Conrad, B.~T.~Fleming, J.~Formaggio, S.~Koutsoliotas,
J.~H.~Kim, C.~McNulty, A.~Romosan, M.~H.~Shaevitz, E.~G.~Stern, A.~Vaitaitis, E.~D.~Zimmerman}
\address{Columbia University, New York, NY, USA}
\author{R.~A.~Johnson, M.~Vakili, N.~Suwonjandee}
\address{University of Cincinnati, Cincinnati, OH, USA}
\author{R.~H.~Bernstein, L.~Bugel, M.~J.~Lamm, W.~Marsh, P.~Nienaber, J.~Yu}
\address{Fermi National Accelerator Laboratory, Batavia, IL, USA}
\author{L.~de~Barbaro, D.~Buchholz, H.~Schellman, G.~P.~Zeller}
\address{Northwestern University, Evanston, IL, USA}
\author{J.~Brau, R.~B.~Drucker, R.~Frey, D.~Mason}
\address{University of Oregon, Eugene, OR, USA}
\author{J. E. McDonald, D. Naples, M. Tzanov}
\address{University of Pittsburgh, Pittsburgh, PA, USA}
\author{S.~Avvakumov, P.~de~Barbaro, A.~Bodek, H.~Budd, D.~A.~Harris,
K.~S.~McFarland, W.~K.~Sakumoto, U.~K.~Yang}
\address{University of Rochester, Rochester, NY, USA}
\date{\today}
\maketitle

\begin{abstract}
We present measurements of the semi-inclusive cross-sections for $\nu _{\mu }$-- 
and $\bar{\nu}_{\mu }$--nucleon deep inelastic scattering interactions with
two oppositely charged muons in the final state. These events dominantly
arise from production of a charm quark during the scattering process. The
measurement was obtained from the analysis of 5102 $\nu _{\mu }$-induced and
1458 $\bar{\nu }_{\mu }$-induced events collected with the NuTeV
detector exposed to a sign-selected beam at the Fermilab Tevatron. We also
extract a cross-section measurement from a re-analysis of 5030 $\nu _{\mu }$-induced 
and 1060 $\bar{\nu }_{\mu }$-induced events collected from the
exposure of the same detector to a quad-triplet beam by the CCFR experiment.
The results are combined to obtain the most statistically precise
measurement of neutrino-induced dimuon production cross-sections to date.
These measurements should be of broad use to phenomenologists interested in
the dynamics of charm production, the strangeness content of the nucleon,
and the CKM matrix element $V_{cd}$.
\end{abstract}

\section{Introduction}

Oppositely-charged dimuon production in neutrino-nucleon deep inelastic
scattering (DIS) provide an excellent source of information on the structure
of the nucleon, the dynamics of heavy quark production, and the values of
several fundamental parameters of the Standard Model (SM) of particle
physics. These events are produced most commonly in charged-current (CC)
neutrino DIS interactions when the incoming neutrino scatters off a strange
or down quark to produce a charm quark in the final state, which
subsequently fragments into a charmed hadron that decays semi-muonically.
This distinct signature is easy to identify and measure in massive
detectors, which allows for the collection of high statistics data samples.
Consequently, dimuon events have played a significant role in the last 20
years in understanding charm production in DIS\cite{rmp}.

Since the contribution of the down quark to charm production is Cabibbo
suppressed, scattering off a strange quark is responsible for a significant
fraction of the total dimuon rate: $\sim 90\%$ in $\bar{\nu}_{\mu }N$ and $%
\sim 50\%$ in $\nu _{\mu }N$ scattering. Therefore the measurement of this
process provides an excellent source for the determination of the strange
quark parton distribution function (PDF) of the nucleon. Other methods
involving measurements of the difference in parity violating structure
functions from $\nu $ and $\bar{\nu }$ scattering off an isoscalar
target and the ratio of parity conserving structure functions from charged
lepton and neutrino DIS suffer from theoretical and experimental systematic
uncertainties. Its prominent role in DIS charm production makes the $s$
quark PDF an important ingredient in tests of two-scale quantum
chromodynamics (QCD)\cite{acot,gkr,mrst,smith,barone}, the two scales being
the squared invariant momentum transfer $Q^{2}$ and the charm mass $m_{c}$.
The $s$ quark PDF also enters into background calculations for new physics
searches at the Tevatron and the LHC. For example, in some scenarios the
super-symmetric top quark $\tilde{t}$ is best searched for via its loop
decay to a charm quark and a neutralino\cite{regina}, $\tilde{t}\rightarrow c%
\tilde{\chi}^{0}$. The dominant background to this mode is from the
gluon-strange quark process $gs\rightarrow W^{-}c$, with the $W^{-}$
decaying leptonically. Accurate separate measurement of $s$ and $\bar{s}$
PDFs may also shed light on the interplay between perturbative and
non-perturbative QCD effects in the nucleon\cite{brodsky}.

In the following we present a new measurement of dimuon production from
scattering of $\nu _{\mu }$ and $\bar{\nu}_{\mu }$ beams from an iron target
in the NuTeV experiment at the Fermilab Tevatron. This measurement exploits
the high purity $\nu _{\mu }$ and $\bar{\nu}_{\mu }$ beams produced by the
Fermilab Sign-Selected Quadrupole Train (SSQT) to extend acceptance for
muons down to $5$ GeV$/c$ momentum without any ambiguity in the 
determination of the muon from the primary interaction.  
Our analysis proceeds as follows: We first perform a fit to
NuTeV $\nu _{\mu }$ and $\bar{\nu}_{\mu }$ dimuon data using a leading order
(LO) QCD cross-section model to obtain values for effective charm mass $%
m_{c} $, parameters describing the size and shape of the nucleon strange and
anti-strange sea, and parameters describing the fragmentation of charm
quarks into hadrons followed by their subsequent semi-muonic decay. The LO
parameters allow us to make contact with previous measurements and provide
an accurate description of the dimuon data. We then go on to use the LO
model for acceptance and resolution corrections in extracting, for the first
time, the cross-sections $d\sigma \left( \nu _{\mu }/\bar{\nu}_{\mu
}Fe\rightarrow \mu ^{\mp }\mu ^{\pm }X\right) /dxdy$ for forward secondary
muons tabulated in bins of neutrino energy $E_{\nu }$, Bjorken scaling
variable $x$, and inelasticity $y$. These cross-section tables provide the
most model-independent convenient representation of $\nu _{\mu }$ and $\bar{%
\nu}_{\mu }$ dimuon data. They may be used to test any cross-section
calculation for dimuon production from iron, provided that the model is
augmented with a fragmentation/decay package such as that provided by PYTHIA 
\cite{pythia}. We then repeat the cross-section extraction on an older CCFR
data set\cite{bazarko} from the same detector; and, after demonstrating
consistency between the CCFR and NuTeV data, we combine the cross-section
tables to obtain the most statistically precise high energy neutrino dimuon
production collected to date. We finish by performing a LO QCD fit to the
combined cross-section tables and extract precise determinations of $m_{c}$,
the CKM parameter $V_{cd}$, and size and shape parameters for the strange
sea.

\section{Leading Order Charm Production}

\begin{figure}[tbp]
\psfig{figure=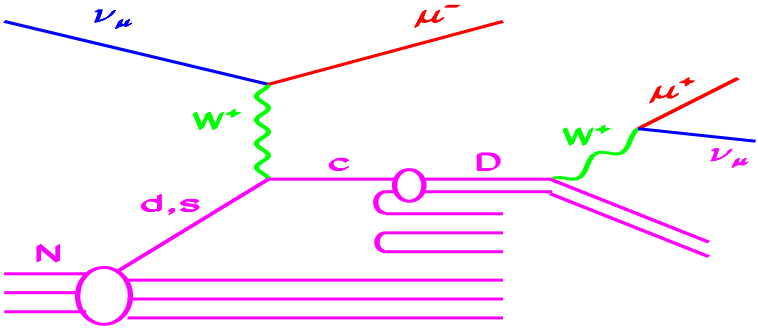,width=0.6\linewidth,height=6cm}
\caption{Dimuon production in $\protect\nu$-nucleon DIS from scattering off
a strange or down quark (LO QCD charm production).}
\label{fig:lodiag}
\end{figure}

Dimuon production from charm depends on three different components: the
charm production cross-section, the fragmentation of the charm quark to a
charmed hadron, and the semileptonic decay of the charmed hadron. In LO QCD,
charm production arises from scattering off a strange or down quark:  
$\nu _\mu \;+\;{\rm N}\;\longrightarrow \;\mu ^{-}\;+\;c\;+\;{\rm X}$,
where the second muon is produced from the semileptonic decay of the 
charmed hadron (Fig.~\ref{fig:lodiag}). In this case, the dimuon production 
cross-section
factorizes into the form: 
\begin{equation}
\text{\hspace*{-0.8cm}}\frac{d^{3}\sigma _{\mu \mu }^{LO}(\nu _{\mu
}N\rightarrow \mu ^{-}\mu ^{+}X)}{dx~dy~dz}=\frac{d^{2}\sigma
_{c}^{LO}(\nu _{\mu }N\rightarrow cX)}{d\xi ~dy}\text{{\ }}D(z)\ B_{c}\text{%
{}}.  \label{lo dimu}
\end{equation}
Here, $y$ is the inelasticity, $z$ is the
fraction of charm quark momentum carried by the charm quark hadron, and $\xi
=x\left( 1+m_{c}^{2}/Q^{2}\right) \left( 1-x^{2}M^{2}/Q^{2}\right) $ is the
fraction of the nucleon's momentum carried by the struck quark, with $x$ the
Bjorken scaling variable and $M$ the nucleon mass. $D(z)$ is
the fragmentation function for the charm quark; $B_{c}$, the semi-muonic
branching fraction for charmed hadrons; both averaged over all
charmed hadrons produced in the final state. The charm production
cross-section $d\sigma (\nu _{\mu }N\rightarrow cX)/d\xi \,dy$ for an
isoscalar target $N$ can be
expressed as

\begin{eqnarray}
\hspace*{-1cm}\hspace*{-0.8cm}\frac{d^{2}\sigma _{c}^{LO}\left( \nu _{\mu
}N\rightarrow \mu ^{-}cX\right) }{d\xi dy} &=&\frac{2G_{F}^{2}ME/\pi }{%
\left( 1+Q^{2}/M_{W}^{2}\right) ^{2}}\left( 1-\frac{m_{c}^{2}}{\ 2ME_{\nu
}\xi }\right)   \label{eq:lo} \\
&&\times \left[ \left| V_{cs}\right| ^{2}s\left( \xi ,Q^{2}\right) +\left|
V_{cd}\right| ^{2}\left( u\left( \xi ,Q^{2}\right) +d\left( \xi
,Q^{2}\right) \right) /2\right] ,  \nonumber
\end{eqnarray}
where $G_{F}=1.17 \times 10^{-5}$ GeV$^{2}$, 
$M_{W}=80.42$ GeV$/c^{2}$, and $%
V_{cs}$/$V_{cd}$ are the $cs$/$cd$ CKM matrix elements. 
The corresponding $\bar{\nu }_\mu $ process has the quarks
replaced by their anti-quark partners.  
One observes that
the non-strange contributions to the cross-section are large in neutrino
mode, where they dominate at high $\xi $; and small in anti-neutrino mode,
where $\bar{s}\left( \xi ,Q^{2}\right) $ dominates at all $\xi $. 
In the case
where $s\left( \xi ,Q^{2}\right) =\bar{s}\left( \xi ,Q^{2}\right) $, the
neutrino $\xi $ distribution determines the relative size of the $s\left(
\xi ,Q^{2}\right) $ and $\left( u\left( \xi ,Q^{2}\right) +d\left( \xi
,Q^{2}\right) \right) /2$ contributions, the anti-neutrino $\xi $
distributions determines the shape of $s\left( \xi ,Q^{2}\right) $, the
energy dependence of the cross-section determines $m_{c}$, the ratio of the
dimuon cross-section to the single muon inclusive cross-section sets $B_{c}$%
, and the energy distribution of the charm decay muon constrains $D\left(
z\right) $.

The non-strange PDFs are determined from inclusive $\nu N$ and $\bar{\nu}N$ CC 
scattering, and from charged lepton scattering; they contribute
negligible uncertainty to $\sigma _{c}$. The final state charmed hadron
admixture is obtained from neutrino data from an emulsion experiment\cite
{e531} that has been corrected for improved knowledge of charmed hadron
lifetimes\cite{e531-tab}. Extraction of CKM matrix elements from the data
requires an independent determination of $B_{c}$; this analysis uses the Particle Data
Group charm semi-leptonic branching fractions\cite{PDG} convolved with the
species production cross-sections just mentioned. The result is $%
B_{c}^{ext}=0.093\pm 0.009$.

In next-to-leading-order (NLO) QCD, additional diagrams complicate the expression for 
$\sigma _{c}$ and also spoil the factorization of the various components of $%
\sigma _{\mu \mu }$. The charm production cross-section also becomes
dependent on the QCD factorization and re-normalization schemes and their
respective scales.

\section{Experimental measurement and analysis technique}

\label{sec1}

In an ideal situation one would like to present direct measurements of the
differential charm production cross-sections $d\sigma _{c}^{\nu \left( \bar{%
\nu}\right) N}/dxdy$ at several different neutrino energies. Neither NuTeV 
nor its
predecessor CCFR  measures charm, but rather dimuons. The charm
cross-section is thus related to the data by model-dependent corrections
for charm fragmentation and decay and by experimental effects of resolution,
acceptance, and neutrino flux. One way of handling these issues is to fit a
parametric model directly to the data and extract parameters from the model.
This approach was used in the past for LO QCD\cite
{rabinowitz,cdhs,charmii,nomad} and NLO QCD\cite{bazarko} in the variable
flavor ACOT\cite{acot} scheme.

The approach taken here begins with the same idea, a LO QCD parametric fit
based on Eq.~\ref{lo dimu}. Events passing selection criteria detailed in
the next section are binned separately in $\nu _{\mu }$ and $\bar{\nu}_{\mu }
$ mode in the quantities 
\begin{eqnarray}
E_{VIS} &=&E_{\mu 1}+E_{\mu 2}+E_{HAD}, \\
x_{VIS} &=&\frac{4E_{\mu 1}\left( E_{\mu 1}+E_{\mu 2}+E_{HAD}\right) \sin
^{2}\theta _{\mu 1}/2}{2M\left( E_{\mu 2}+E_{HAD}\right) }, {\rm and}\\
z_{VIS} &=&\frac{E_{\mu 2}}{E_{\mu 2}+E_{HAD}},
\end{eqnarray}
where $E_{\mu 1}$ is the energy of the primary muon with the same lepton
number as the beam, $E_{\mu 2}$ is the energy of the other muon, $E_{HAD}$
is the observed hadronic energy in the calorimeter, and $\theta _{\mu 1}$ is
the scattering angle of the primary muon. The ``$VIS$'' subscript indicates
that these quantities differ from the true values of $x$, $E$, and $z$, due
to the energy carried away by the neutrino from charm decay and due to
detector smearing. Other quantities of interest for comparison purposes are 
\begin{eqnarray}
y_{VIS} &=&\frac{E_{\mu 2}+E_{HAD}}{E_{\mu 1}+E_{\mu 2}+E_{HAD}}~~~~{\rm and} \\
Q_{VIS}^{2} &=&2ME_{VIS}x_{VIS}y_{VIS}.
\end{eqnarray}

A binned likelihood fit is performed which compares the data to a model
composed of a charm source described by Eqs.~\ref{lo dimu} and \ref{eq:lo}.
The charm events are augmented with a contribution from dimuon production
through $\pi /K$ decay in the charged-current neutrino interaction's hadron
shower and then processed through a detailed Monte Carlo (MC) simulation of
the detector and the same event reconstruction software used for the data.
The MC dimuon sample is normalized to the data through use of the inclusive
single muon event rates in $\nu _{\mu }$ and $\bar{\nu}_{\mu }$ mode. The
fit varies a common charm mass $m_{c}$, branching fraction $B_{c}$, and
fragmentation parameter $\epsilon $ for both modes, and two parameters for
each mode, $\left( \kappa _{\nu },\alpha _{\nu }\right) $ and $\left( \kappa
_{\bar{\nu}},\alpha _{\bar{\nu}}\right) $, that describe the magnitude and
shape of the $s$ and $\bar{s}$ quark PDFs. The strange sea parameters are
defined by 
\begin{eqnarray}
s\left( x,Q^{2}\right) &=&\kappa _{\nu }\frac{\bar{u}\left( x,Q^{2}\right) +%
\bar{d}\left( x,Q^{2}\right) }{2}\left( 1-x\right) ^{\alpha _{\nu }}~~~ {\rm and}
\label{ssconstr} \\
\bar{s}\left( x,Q^{2}\right) &=&\kappa _{\bar{\nu}}\frac{\bar{u}\left(
x,Q^{2}\right) +\bar{d}\left( x,Q^{2}\right) }{2}\left( 1-x\right) ^{\alpha
_{\bar{\nu}}}.  \label{sbsconstr}
\end{eqnarray}
\renewcommand{\thefootnote}{\fnsymbol{footnote}}
In these parameterizations\footnote{%
This parameterization differs slightly from previous LO analyses in the definition of $%
\kappa _{\nu }$ and $\kappa _{\bar{\nu}}$ in the general case of $\alpha _{\nu
}$ and $\alpha _{\bar{\nu}}\neq 0$. The motivation for not using the older
definitions (of the form $s\left( x,Q^{2}\right) =\frac{\kappa _{\nu }}{2}%
\left( \bar{u}\left( x,Q^{2}\right) +\bar{d}\left( x,Q^{2}\right) \right)
\left( 1-x\right) ^{\alpha _{\nu }}$ $\times $ $\frac{\int_{0}^{1}dx\left( 
\bar{u}\left( x,Q^{2}\right) +\bar{d}\left( x,Q^{2}\right) \right) }{%
\int_{0}^{1}dx\left( \bar{u}\left( x,Q^{2}\right) +\bar{d}\left(
x,Q^{2}\right) \right) \left( 1-x\right) ^{\alpha _{\nu }}}$) is to avoid a
procedure that requires information about the PDF outside the experimentally
accessible $x$ range of the experiment.}, values of $\kappa _{\nu }=\kappa _{%
\bar{\nu}}=1$ and $\alpha _{\nu }=\alpha _{\bar{\nu}}=0$ would imply an
SU(3)-flavor symmetric sea; previous measurements have yielded $\kappa $
values around $0.4$, and $\alpha $ values consistent with zero 
(within large errors). The fragmentation process is described using the
Collins-Spiller fragmentation function~\cite{collins}: 
\begin{equation}
D(z,\epsilon )=N\left[ (1-z)/z+\epsilon _{C}(2-z)/(1-z)\right]
(1+z)^{2}[1-(1/z)-\epsilon _{C}/(1-z)]^{-2}.
\end{equation}
The Peterson function~\cite{peterson} was also tried but produced worse
agreement between MC and data.

The analysis proceeds based on the observation that the dimuon data are
well described by the LO fit. The information on charm production 
from the LO fit is,   
in fact, not of great importance; our goal is to use the fit
parameters to help construct a cross-section. This task is performed by
forming a new grid in $x_{VIS}$, $y_{VIS}$, and $E_{VIS}$ in the data and
MC, and a corresponding grid of $x,y,$ and $E$ in the MC. The dimuon 
cross-section is computed at the weighted center of each $\left( x,y,E\right) $
bin $i$. The MC also predicts, with the result of the LO fit, the number of
events in each $\left( x_{VIS},y_{VIS},E_{VIS}\right) $ bin $j$. The MC can
further be used to establish a correspondence between $\left( x,y,E\right) $
and $\left( x_{VIS},y_{VIS},E_{VIS}\right) $ bins; this is accomplished by
finding the bin $j_{i}$ in $\left( x_{VIS},y_{VIS},E_{VIS}\right) $ space
that receives the largest fraction of events produced in $\left(
x,y,E\right) $ bin $i$. After this procedure the cross-section for $\left(
x,y,E\right) $ bin $i$ is then determined by 
\begin{equation}
\hspace*{-0.8cm}\left\{ \frac{d^{2}\sigma _{\mu \mu }(\nu _{\mu }N\rightarrow
\mu ^{-}\mu ^{+}X)}{dx~dy\ }\right\} _{\left( x,y,E\right) _{i}}{=}\frac{%
D_{i}}{{\cal N}_{i,\text{LO fit}}^{\prime }}\int_{E_{\mu 2}>E_{\mu 2\min
}}dz~d\Omega \left\{ \frac{d^{3}\sigma _{\mu \mu }(\nu _{\mu }N\rightarrow
\mu ^{-}\mu ^{+}X)}{dx~dy~dz}\right\} _{\left( x,y,E\right) _{i}-\text{LO
fit}}.  \label{measured XC}
\end{equation}
In this expression the left hand side represents the measured cross-section
for dimuon production as a function of $x$, $y$, and $E$ in bin $i$ with the
requirement that the second muon in the event exceeds the threshold used in
the experiment, $E_{\mu 2}>E_{\mu 2\min }$. On the right hand side, $D_{i}$
is the number of data corresponding to bin $i$, and ${\cal N}_{i,\text{LO fit}%
}^{\prime }$ is the corresponding number of events predicted by the LO fit.
In the integrand, $d^{3}\sigma _{\mu \mu }(\nu _{\mu }N\rightarrow \mu
^{-}\mu ^{+}X)/dx\ dy\ dz$ is taken from Eq.~\ref{lo dimu}, and the integral over
the fragmentation variable $z$ and charmed hadron decay variable $\Omega $
maintains the condition $E_{\mu 2}>E_{\mu 2\min }$. The procedure for
defining $D_{i}$ and 
${\cal N}_{i,\text{LO fit}}^{\prime }$ is detailed further in
Appendix \ref{binstuff}.

The end result is two tables, one each for $\nu _{\mu }$ and $\bar{\nu}_{\mu
}$ mode, of the ``forward'' dimuon cross-section (which is closest to what
the experiment actually measures). It will be shown later that these tables
can be used to re-extract the LO fit parameters and can be combined with
similar tables from the CCFR experiment. More details on the apparatus,
event selection, and analysis procedure will be given first.

\section{NuTeV Experimental Details}

\subsection{Detector and beamline}

\begin{figure}[tbp]
\psfig{file=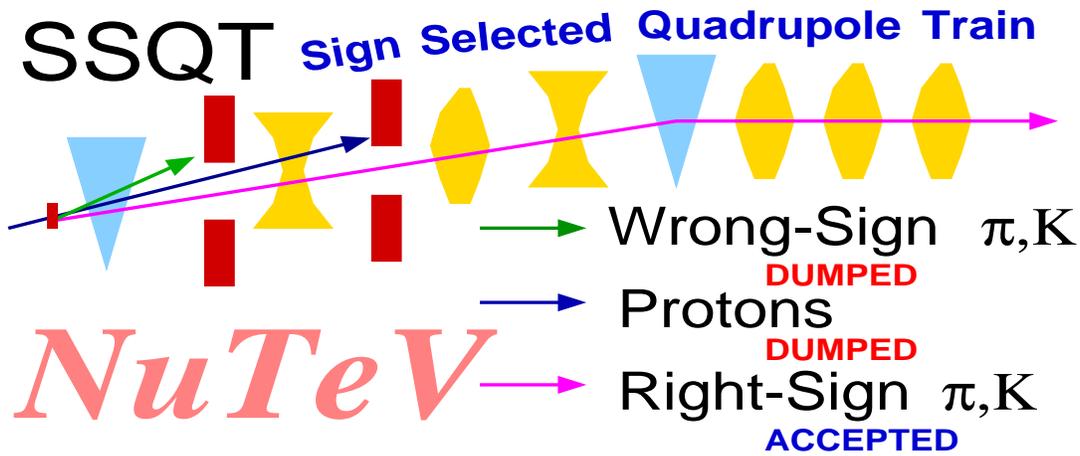,width=0.8\linewidth,height=6cm}
\caption{Schematic of the NuTeV beamline.}
\label{fig:ssqt}
\end{figure}

The NuTeV (Fermilab-E815) neutrino experiment collected data during 1996-97
with the refurbished Lab E neutrino detector and a newly installed
Sign-Selected Quadrupole Train(SSQT) neutrino beamline. Figure \ref{fig:ssqt}
illustrates the sign-selection optics employed by the SSQT to pick the
charge of secondary pions and kaons which determine whether $\nu _{\mu }$ or 
$\bar{\nu}_{\mu }$ are predominantly produced. The SSQT produced beam
impurities of $\bar{\nu }_{\mu }$ ($\nu _{\mu }$) events in 
$\nu _{\mu }$ mode ($\bar{\nu }_{\mu }$ mode) at the $10^{-3}$ level. During NuTeV's
run the primary production target received $1.13\times 10^{18}$ and $%
1.41\times 10^{18}$ protons on target in neutrino and anti-neutrino modes,
respectively, resulting in inclusive CC samples of $1.3\times 10^{6}$ events
in neutrino mode and $0.46\times 10^{6}$ $\bar{\nu }$ inclusive CC
events. The very low ``wrong-flavor'' backgrounds\cite{drew} imply that only
one muon charge measurement is needed to make the correct assignments for $%
E_{\mu 1}$ and $E_{\mu 2}$ described above.

The Lab E detector, described in detail elsewhere\cite{nim}, consists of two
major parts, a target calorimeter and an iron toroid spectrometer. The
target calorimeter contains 690 tons of steel sampled at 10~cm(Fe) intervals
by 84 3m $\times$ 3m scintillator counters and at 20~cm(Fe) intervals by 
42 3m $\times$ 3m drift chambers. The toroid spectrometer consists of four
stations of drift chambers separated by three iron toroid magnets that
provide a $p_{T}$ kick of $2.4$ GeV$/c$. The toroid magnets were set to
always focus the muon with same lepton number as the beam neutrino.
Precision hadron and muon calibration beams monitored the calorimeter and
spectrometer performance throughout the course of data taking. The
calorimeter achieves a sampling-dominated resolution of $\sigma
_{E}/E=2.4\%\oplus 87\%/\sqrt{E}$ and an absolute scale uncertainty of $%
\delta E/E=0.4\%$. The spectrometer's multiple-Coulomb-scattering-dominated
muon momentum resolution is $\sigma _{p}/p=$ $11\%$, and the muon momentum
scale is known to $\delta p/p=1.0\%$.

\subsection{Data selection} \label{dataselection}

\begin{figure}[tbp]
\centering{
\psfig{file=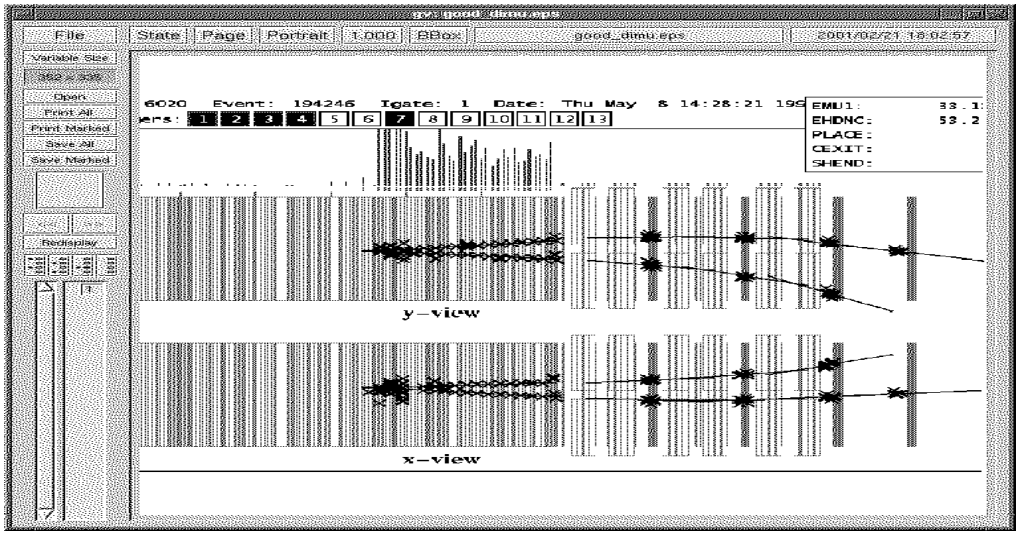,width=0.75\linewidth,height=7cm}
}
\caption{Typical dimuon event.}
\label{fig:evt}
\end{figure}

A typical dimuon event has the characteristics shown in Fig.~\ref{fig:evt}.
In this figure, the toroid can be seen to focus the leading muon originating
from the leptonic vertex and to de-focus the secondary muon, which originates
most probably from charm decay. In the event shown both muons pass through
the toroid, and both their signs are measured. In events where the sign of
one muon is not measured, it is assumed to be opposite the one
measured. Since the sign of the primary muon is known because of the sign
selection of the SSQT, the measurement of the sign of only one muon is
sufficient to identify the primary and secondary muon in the event. The
rate of the same sign dimuon events with both muons toroid analyzed is very
low\cite{drew}.

Candidate opposite sign dimuon events, in both data and Monte Carlo,
were selected using the following criteria:

\begin{itemize}
\item  {The event must occur in coincidence with the beam and fire the
penetration trigger (charged-current interaction trigger).}

\item  {The incident neutrino energy, $E_{\nu }$, must be greater than }${20}
${\ GeV and the energy of the hadronic shower, $E_{had}$, greater than }${10}
${\ GeV.}

\item  {In order to ensure event containment, only events occurring within
an active fiducial volume are accepted: the transverse vertex positions }$%
\left( V_{x},V_{y}\right) ${\ must be satisfy $-127$ cm$<V_{x,y}<127$ cm and 
$\sqrt{V_{x}^{2}+V_{y}^{2}}<152.4$ cm, and the longitudinal vertex
position must lie between counters 15 and 80, which corresponds to }$2.7$
and $13.3$ hadronic interaction lengths from the upstream and downstream
ends of the calorimeter, respectively.

\item  {Two muons must be identified in the event and satisfy the following
criteria:}

\begin{itemize}
\item  {The energy of each of the two muons, $E_{\mu 1,\mu 2}$, must be greater than 5
GeV.}

\item  {The time obtained from fitting each track must be within 36 ns of
the trigger time. }

\item  {One of the muons must be toroid-analyzed and the energy of the
toroid-analyzed muon at the entrance of the toroid must be$\ $greater than $%
3 $ GeV.}

\item  {At least one toroid-analyzed muon must pass through at least $2/3$
of the toroid.}

\item  {The toroid-analyzed muons must hit the front face of the toroid
inside a circle of radius $R_{FF}<152$~cm, and more than $80\%$ of the
path length of the muon must be in the toroid steel.}
\end{itemize}

\item  {Finally, in order to remove mis-reconstructed events, a requirement
is imposed on the reconstructed $x_{VIS}$ kinematic variable: $0\leq
x_{VIS}\leq 1$.}
\end{itemize}

The final event sample contains $5102$ $\nu _{\mu }$-induced and $1458$ $%
\bar{\nu }_{\mu }$-induced events. Of these, $2280/655$ in $\nu _{\mu }/%
\bar{\nu }_{\mu }$ mode have both muons reconstructed in the toroid
spectrometer.  All other events have only one. The mean $E_{VIS}$ of the
events is $157.8$ GeV, the mean $Q_{VIS}^{2}=21.1$~GeV$^{2}$, and the mean $%
x_{VIS}=0.14$.  The overall reconstruction efficiency, including the detector 
acceptance, is $\sim 60\%$ for events with $E_{\mu 2} \approx$ 5 GeV, and 
$\sim 80\%$ when  $E_{\mu 2}$ is above 30 GeV.

\subsection{Detector Simulation}

\label{mc}

A hit-level Monte Carlo simulation of the detector based on the GEANT
package~\cite{geant} was used to model the detector response and provide an
accurate representation of the detector geometry. The Monte Carlo events
were analyzed using the same reconstruction software used in the data
analysis. The detector response in the simulation was tuned to both hadron
and muon test beam data at various energies. To ensure the accurate modeling
of the muon reconstruction efficiency the drift chamber efficiencies
implemented in the simulation were measured in the data as a function of
time.

The primary neutrino interactions were generated using the LO QCD model and
fragmentation function described in Sec.~\ref{sec1}. Electroweak radiative
corrections based on the model by Bardin~\cite{bardin} were applied to this
cross-section. The main background source from ordinary CC interactions in
which a pion or kaon produced in the hadronic shower decays muonically was
generated following a parameterization of hadron test beam muoproduction
data for simulating secondary decays, and the LEPTO~\cite{leplu} package for
the decays of primary hadrons~\cite{pams}.  The total probability to produce
such muons with momentum greater than 4 GeV/c is $\approx 2\times10^{-4}$
for events with $E_{had} \sim $ 30 GeV, and $\approx 10^{-3}$ for 
$E_{had} \sim$ 100 GeV.  The contribution from the primary hadron decays is 
roughly two times larger than that from the secondary decays.

\subsection{Neutrino Flux and Normalization}

The total flux, energy spectra, and composition for both neutrino and
anti-neutrino beams are calculated using a Monte Carlo simulation of the
beamline based on the DECAY TURTLE program~\cite{turtle} and production data
from Atherton~\cite{atherton} as parameterized by Malensek~\cite{malensek}
for thick targets. This flux is used to generate an inclusive
charged-current interaction Monte Carlo sample using the GEANT based 
hit-level detector Monte Carlo described in section IV.C.

The predicted flux is then tuned so that the inclusive charged-current
interaction spectra in the Monte Carlo match the data. Selection criteria
for this sample of inclusive charged-current interactions are exactly the
same as those used to select the dimuon sample with the requirement for two
muons removed. Flux corrections of up to $15\%$ are applied in bins of
neutrino energy and transverse vertex position to force the single muon data
and Monte Carlo to agree. In addition, an overall factor is determined for
each beam (neutrino or anti-neutrino) that absolutely normalizes the single
muon Monte Carlo to the data. The dimuon Monte Carlo uses the flux
determined with the above procedure; and it is absolutely normalized to the
inclusive single muon charged-current data through the flux tuning procedure.

The procedure used to tune the flux to the observed single muon rate is
iterative since the event rate observed in the detector depends on the
convolution of the cross-section with neutrino flux, and the result of the
charm measurement has a small effect on the total cross-section. Corrections
found from the single muon Monte Carlo/data comparisons are used in the
dimuon Monte Carlo that is used to determine the dimuon cross-section, and
thus the charm production cross-section within our LO model. The charm
cross-section results are then used in the single muon Monte Carlo and the
procedure is repeated until the flux parameters do not change (in practice
the convergence is very fast). Figure \ref{fig:1mupl} shows a comparison
between data and Monte Carlo for the single muon (flux) sample; the level of
agreement is very good. Note that  muon and hadron energies are not adjusted
separately in the flux tuning procedure.

\section{Results}

\subsection{NuTeV and CCFR Leading Order QCD fits}

\begin{figure}[tbp]
\centering{
\psfig{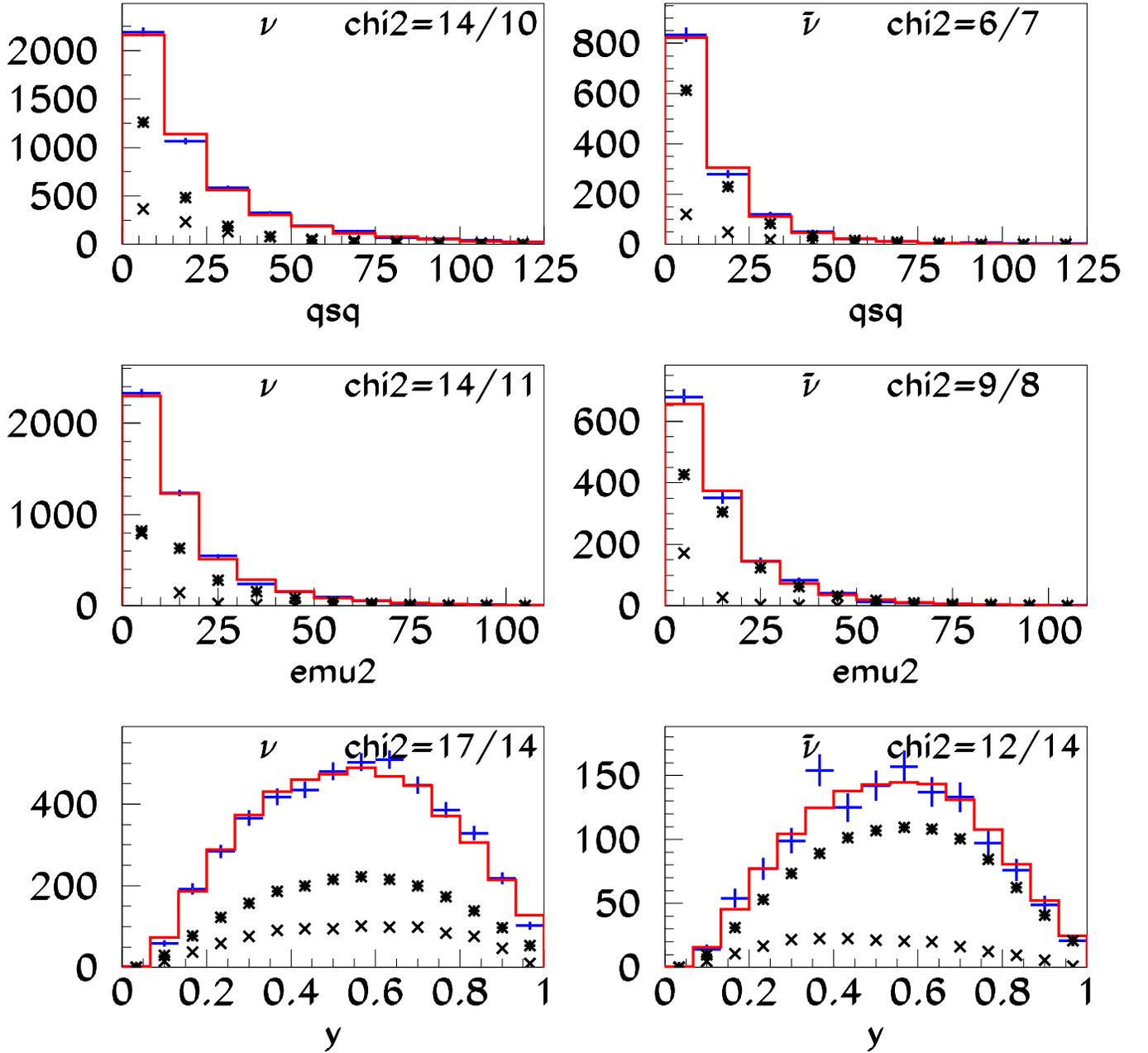}
}
\caption{Dimuon data kinematic distributions (points) 
not used in the fit: $Q^2$ in
GeV$^2$, $E_{\mu 2}$ in GeV, and $y$, from top to bottom.  Neutrinos are  
shown on the left, anti-neutrinos on the right.  The total Monte Carlo 
prediction is represented by the histogram, while the crosses show the 
$\protect\pi K$ background component, and the stars the strange sea
contribution. }
\label{fig:dimuplt}
\end{figure}

The LO QCD fits were performed using three different parton distribution
function (PDF) sets with their corresponding QCD evolution kernels: GRV94LO 
\cite{grv94} and CTEQ4LO\cite{cteq5} , as implemented in the PDF compilation
PDFLIB~\cite{pdflib}, and a Buras-Gaemers parameterization\cite{bgpar}
(BGPAR) that has been used extensively in this experiment and its CCFR
predecessor. In the BGPAR case, an explicit Callan-Gross relation violation
is implemented by replacing the term $1-\frac{m_{c}^{2}}{2ME_{\nu }\xi }$ in
Eq.~\ref{eq:lo} with $(1+R_{L})(1+(\frac{2M\xi }{Q})^{2})^{-1}(1-y-%
\frac{Mxy}{2E})+\frac{xy}{\xi }$, where $R_{L}$, the ratio of longitudinal
to transverse $W^{\pm }N$ cross-sections, is taken from a fit to
electroproduction data\cite{whitlow}. Table \ref{LO data fits} lists fit
results with the rightmost column showing the combined $\chi ^{2}$ for $\nu $
and $\bar{\nu}$ modes. All three models have the same good level of
agreement with the dimuon data. Figure \ref{fig:dimufit} illustrates the
quality of the BGPAR fit by comparing it to the data for
the kinematic variables used directly in the fit; figure \ref{fig:dimuplt}
shows a comparison for variables not used directly in the fit. One
observes that differences in the choice of PDF parameterization result in
different charm production parameters, indicating significant model
dependence at LO. Because of this dependence, the most relevant quantity to
extract is the dimuon production cross-section.

\subsection{CCFR Leading Order QCD fits}

The CCFR dimuon data set used in the CCFR LO~\cite{rabinowitz} and NLO~\cite
{bazarko} analysis is used to extract LO QCD parameters using the same
procedure described in the previous section. Results, shown in Table~\ref
{set data fits}, are consistent with the NuTeV fits except for the
fragmentation function shape parameter $\epsilon $. This difference is
caused by the fact that in NuTeV a much higher percentage of low energy
muons is used, thus the $z_{VIS}$ distribution has significant shape
differences to that of CCFR. In addition, the uncertainty in the $\pi /K$
background parameterization is larger for low muon energies; this effect is
reflected in the size of the error on the determination of $\epsilon $.

Table \ref{set data fits} shows results of a combined fit to the data from
both NuTeV and CCFR performed using the same procedure and the BGPAR PDF
set. Since the fragmentation shape parameter is different for the two
experiments, the Monte Carlo sample for each is reweighted to the
appropriate $\epsilon $ and then kept fixed in the combined fit to simplify
the fitting procedure.

\subsection{Comparing CCFR with NuTeV}

Different BGPAR PDF sets are used to analyze NuTeV and CCFR data. Thus,
since we have shown that different PDF sets can produce different LO
parameters, the numbers presented in Table~\ref{set data fits} should not be
compared directly with each other. Furthermore, the combined fit uses the
NuTeV BGPAR PDF set, an approach that is not completely rigorous. Parameters in Table~\ref
{set data fits} should be compared to their corresponding values in Table~{%
\ref{set table fits},} which gives the result of the fit to the 
dimuon cross-section table discussed in the next section. In this case, the use
of any PDF set or model is equally appropriate, since the purpose is
to compare the results of the fit to the dimuon data to those of the fit to
the extracted cross-section tables.

We caution readers against taking $\kappa $ and $\alpha $ parameters
for strange seas extracted using the BGPAR sets and using them to construct $%
s\left( x,Q^{2}\right) /\bar{s}\left( x,Q^{2}\right) $ PDFs according to
Eqs.~\ref{ssconstr}-\ref{sbsconstr} from a different PDF set. Since the
NuTeV/CCFR PDFs are not publicly available, a safer course would be to take
the strange sea parameters for CTEQ or GRV. Even then, it is important to
match $\kappa $ and $\alpha $ from our fits with the appropriate PDF set.

\subsection{Systematic uncertainties}

Although the main thrust of the analysis is the extraction of the dimuon
production cross-section, the various sources of systematic uncertainty are
presented by listing their contributions to the LO fit parameters. This is
done for reasons of clarity, since the individual systematic uncertainty
contributions add too many entries in the cross-section tables.  These methods are
completely equivalent since the systematic uncertainty on the parameters of
the LO fits propagates directly to the cross-section measurement.

The main sources of the systematic uncertainties arise from modeling
uncertainties in the Monte Carlo. The most significant are the $\pi /K$
decay background simulation, the detector calibration from the analysis of
test beam hadron and muon data as a functions of energy and position, and
the overall normalization. In addition, in the case of the BGPAR fits, the
uncertainty on the longitudinal structure function is important.

The systematic error sources are given in Table~\ref{lo systematics}. For
the combined NuTeV+CCFR fit, systematic errors begin to dominate the
uncertainties for several fit parameters.

\subsection{NuTeV Cross-Section Tables}

As was shown in the previous section,  charm cross-section parameters depend on
the details of the charm production model used in the fit. In addition, a
fragmentation and decay model must be used to extract the charm production
cross-section from the observed dimuon rate, introducing further model
dependence. These model dependences are exacerbated by the experimental
smearing correction, due to the missing neutrino energy of the charmed
hadron decay and the detector resolution, and the substantial acceptance
correction for low energy muons. In order to minimize model dependencies, we
choose to present our result in the form of a dimuon production
cross-section.

We have shown in the previous section that we can obtain very good
description of our dimuon data, independent of the charm production model
assumptions in our Monte Carlo. We thus limit the use of the Monte Carlo to
correct experimental effects to the measured dimuon rate, and we limit the
measurement of this rate to regions of phase space where the acceptance of
the experiment is high. In this case, the only model dependence comes from
potential smearing effects close to our acceptance cuts, i.e., the
uncertainty on production of events produced with kinematic variables
outside our cuts that smeared to reconstructed values within the cuts. The
prediction for this kind of smearing depends on the underlying physics
model, which is not well constrained by our data, since it involves phase
space not accessed by our data. However, this model dependence is a second
order effect.

\begin{figure}[tbp]
\psfig{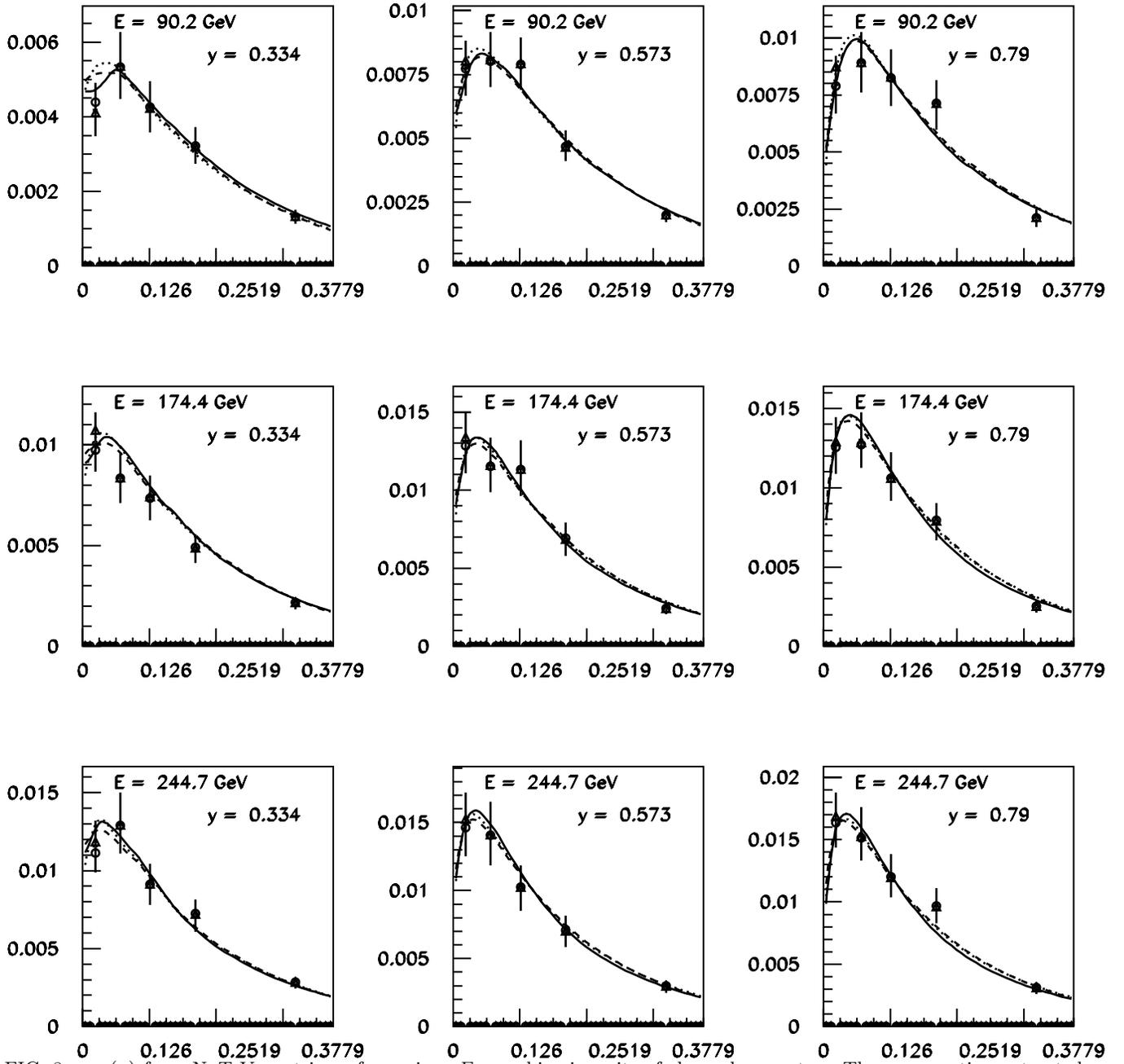}
\caption{$\protect\sigma_{2\protect\mu}(x)$ from NuTeV neutrinos, 
for various $E_\protect\protect%
\nu-y$ bins in units of charged-current $\protect\sigma$.  The cross-section 
extracted using the BGPAR model in the Monte Carlo is shown in squares, the
circles correspond to extraction using the CTEQ model, and the triangles to 
the GRV model.  The curves show the model prediction for GRV (dashed), CTEQ
(dotted), and BGPAR (solid) after the models have been fit to
the data. }
\label{nttytab1}
\end{figure}

\begin{figure}[tbp]
\psfig{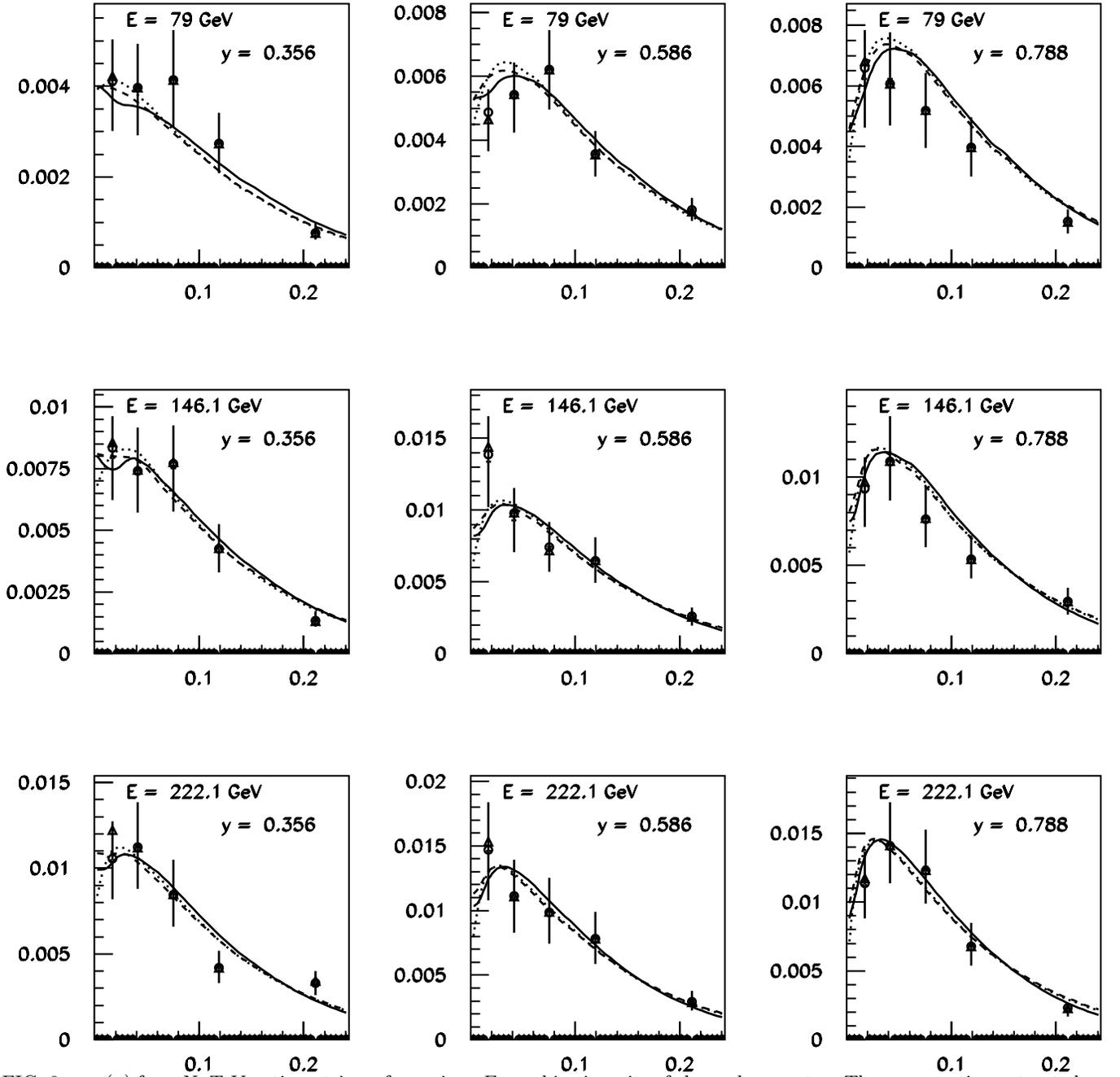}
\caption{$\protect\sigma_{2\protect\mu}(x)$ from NuTeV anti-neutrinos, 
for various $E_\protect\protect%
\nu-y$ bins in units of charged-current $\protect\sigma$.  The cross-section 
extracted using the BGPAR model in the Monte Carlo is shown in squares, the
circles correspond to extraction using the CTEQ model, and the triangles to 
the GRV model.  The curves show the model prediction for GRV (dashed), CTEQ
(dotted), and BGPAR (solid) after the models have been fit to
the data. }
\label{nttytab2}
%
\end{figure}

\begin{figure}[tbp]
\psfig{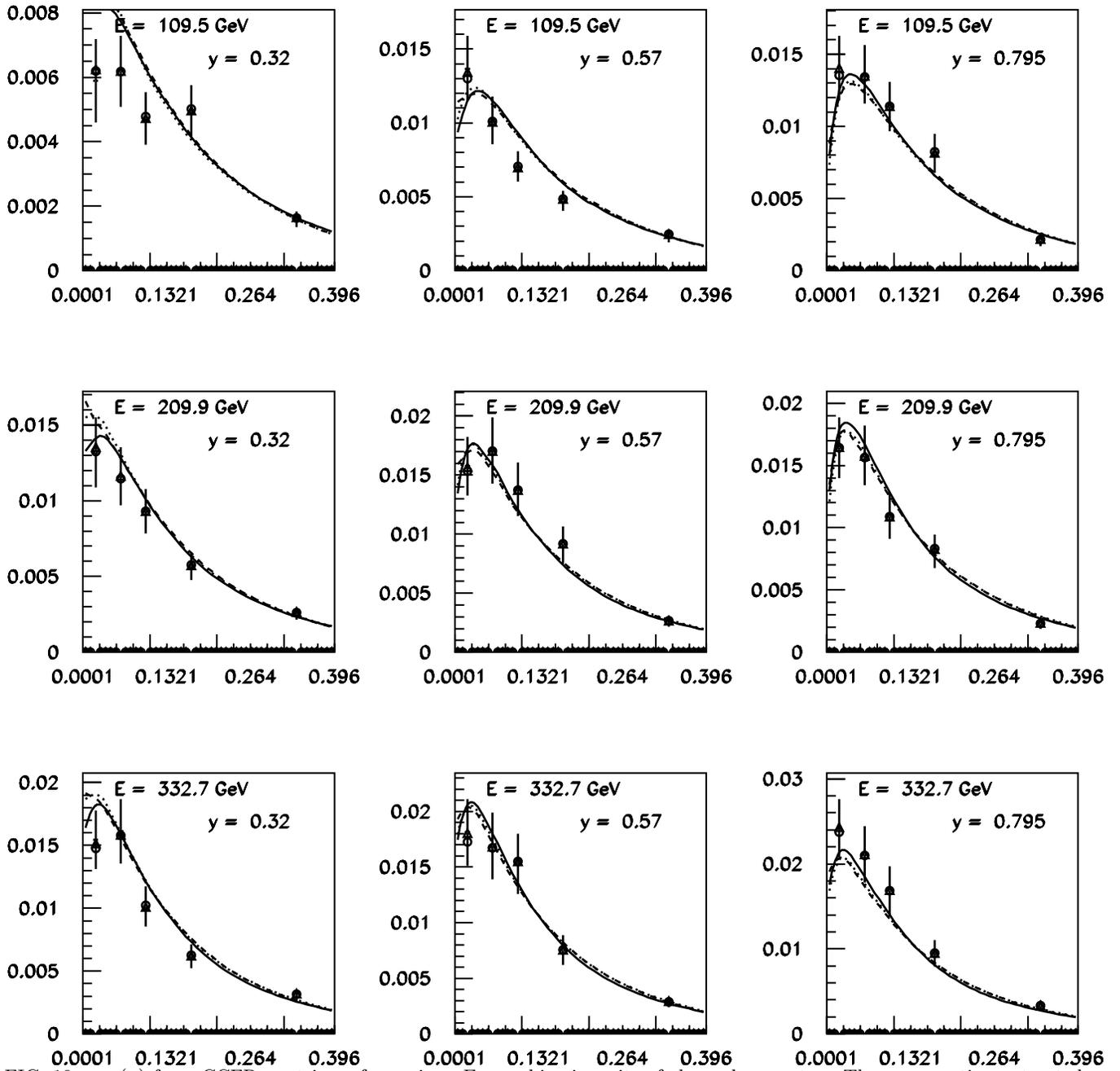}
\caption{$\protect\sigma_{2\protect\mu}(x)$ from CCFR neutrinos, 
for various $E_\protect\protect%
\nu-y$ bins in units of charged-current $\protect\sigma$.  The cross-section 
extracted using the BGPAR model in the Monte Carlo is shown in squares, the
circles correspond to extraction using the CTEQ model, and the triangles to 
the GRV model.  The curves show the model prediction for GRV (dashed), CTEQ
(dotted), and BGPAR (solid) after the models have been fit to
the data. }
\label{nttytab3}
%
\end{figure}

\begin{figure}[tbp]
\psfig{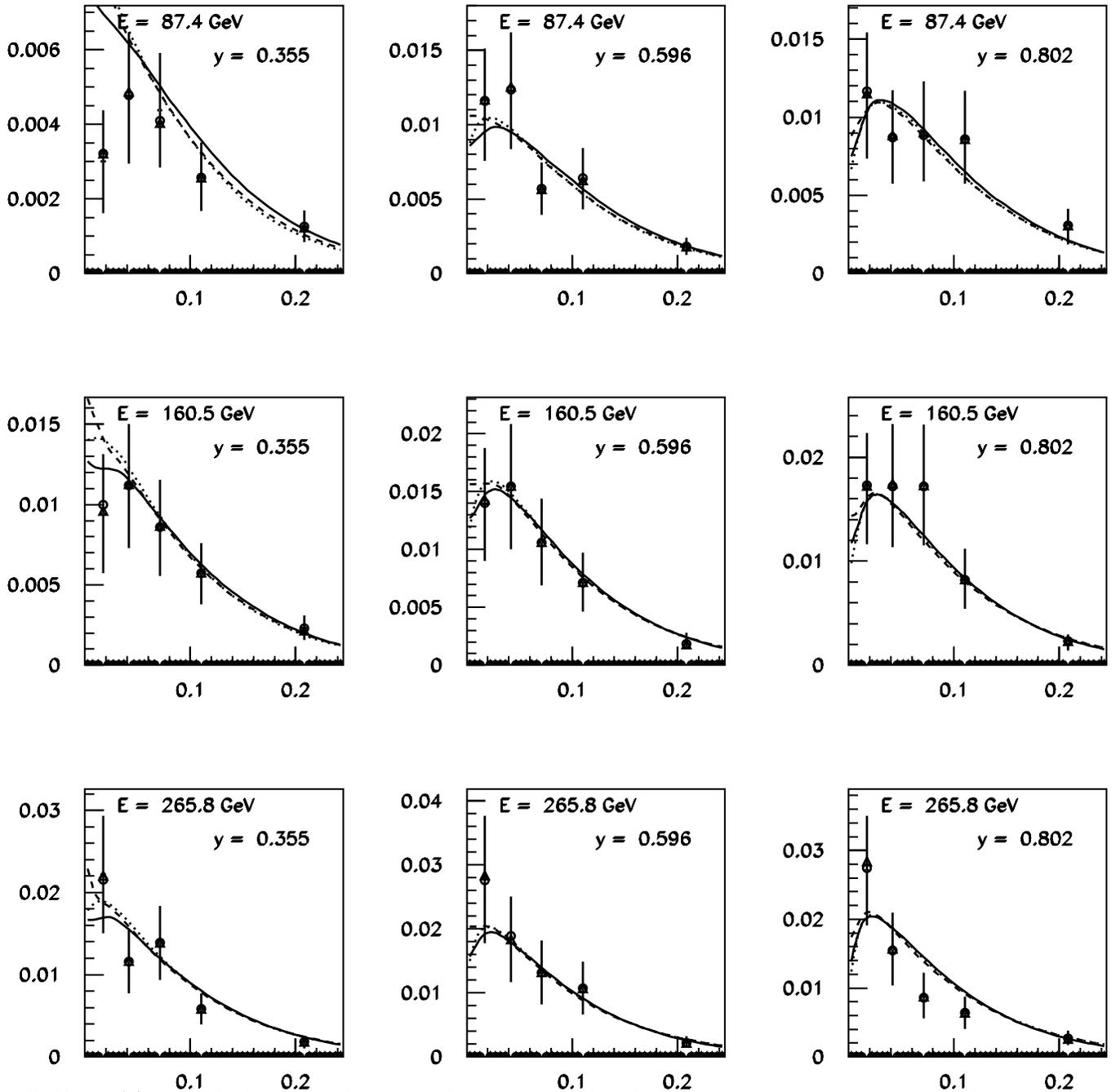}
\caption{$\protect\sigma_{2\protect\mu}(x)$ from CCFR anti-neutrinos, 
for various $E_\protect\protect%
\nu-y$ bins in units of charged-current $\protect\sigma$.  The cross-section 
extracted using the BGPAR model in the Monte Carlo is shown in squares, the
circles correspond to extraction using the CTEQ model, and the triangles to 
the GRV model.  The curves show the model prediction for GRV (dashed), CTEQ
(dotted), and BGPAR (solid) after the models have been fit to
the data. }
\label{nttytab4}
\end{figure}

The dimuon cross-section extraction procedure depends on the ability of our
Monte Carlo to describe the data. The ``true'' three-dimensional phase space 
$\left( x,y,E\right) $ is divided into a number of bins. The grid is set up
so that there is the same number of Monte Carlo events in each bin for any
projection into one dimension. The cross-section for dimuon production with $%
E_{\mu 2}\geq 5$ GeV is calculated according to Eq.~\ref{measured XC} at the
center of gravity of each bin. The size of the bins in ``visible'' phase space $%
\left( x_{VIS},y_{VIS},E_{VIS}\right) $ is defined in such a way so that
there is a correspondence between a visible bin $j_{i}$ and a generated $i$
established by the condition that bin $j_{i}$ contains at least $60\%$ of
the events from bin $i$. The relative error on the cross-section is taken
to be equal to $\delta _{\sigma _{2}}/\sigma _{2\mu }=\sigma _{2\mu }/\sqrt{%
D\left( j_{i}\right) }$. Cross-section tables were produced using all three
BGPAR, GRV, and CTEQ models; figure \ref{nttytab1}-\ref{nttytab4} 
demonstrate the insensitivity
of the cross-section to PDF choice. Further details of the binning
procedure are defined in Appendix \ref{binstuff}.

\subsection{Charm Production Fits Using the Dimuon Cross-Section Tables}

Using the cross-section tables involves similar steps as used in the direct
data analysis, but with all detector and flux dependent effects removed. One
must provide a model for charm production, fragmentation, and decay;
construct the dimuon cross-section number for each entry in the table;
and perform a $\chi ^{2}$ fit. The $\chi ^{2}$ function should employ the
statistical and systematic errors in each bin added in quadrature.

One must also account for correlations between the various table entries.
These correlations derive from our use of the LO fit to parameterize the
data and from our method of binning; they are an inherent consequence of the
incomplete kinematic reconstruction of the dimuon final state. We have
adopted a pragmatic approach towards handling this issue. Rather than
compute a large correlation matrix, we inflate the cross-section errors
in each bin by a factor that is typically $1.4.$ This factor is chosen so
that an {\em uncorrelated} $\chi ^{2}$ fit to the tables returns the same
parameter errors as a direct fit to the data. A consequence of this error
inflation is an apparent best fit $\chi ^{2}$ per bin $\left( \chi ^{2}/%
\text{bin}\right) $ that is approximately $1/4$ rather than $1$. This low
value results from overcounting the number of degrees of freedom (DOF) 
present in the table and is compensated
for by using an effective DOF per bin determined via MC\
calculation and given along with cross section data in the tables.
Further details of the determination of the effective DOF 
are presented in Appendix \ref{binstuff}.

We tested this fitting procedure on the tables using the same BGPAR,
fragmentation, and decay models used to obtain the table; Table~\ref{set
table fits} summarizes this study. Both the parameter values and their
uncertainties obtained from fitting to the cross-section table agree with
the corresponding values obtained by fitting directly to the data. It has
also been verified that GRV94 parameters, for example, can be extracted from
a cross-section table constructed with either the CTEQ or BGPAR model so as
to agree with parameters obtained by fitting directly to the data.

While our cross-checks in fitting the cross-section tables entail using the
same physics model used in generating the tables, we emphasize that the
table presents a set of physical observables which may be used to test {\em %
any} dimuon production model. For the most interesting case of dimuon
production through charm, a typical model test would consist of the
following steps\footnote{%
A simple PYTHIA implementation of the first three steps may be obtained at
www-e815.fnal.gov, or by contacting the authors.}:

\begin{itemize}
\item  For each bin, generate a (large) ensemble of $N_{GEN}$ events with $x$%
, $y$, and $\vec{p}_{c}$, where $\vec{p}_{c}$ denotes the lab momentum of
the produced charm quark, according to the model charm production
differential cross-section $d\sigma _{c-%
\mathop{\rm mod}%
}/dx~dy~d\vec{p}_{c}$

\item  Fragment the charm quarks into hadrons and decay the charmed hadrons
(using, for example, PYTHIA), and determine $N_{PASS}$ the number of events
which have a charm decay muon with $E_{\mu 2}\geq 5$ GeV.

\item  The cross-section to compare to the table value is then $d\sigma
_{\mu \mu -%
\mathop{\rm mod}%
}/dx~dy=\left( N_{PASS}/N_{GEN}\right) \times \int d\vec{p}_{c}~d\sigma _{c-%
\mathop{\rm mod}%
}/dx~dy~d\vec{p}_{c}$.

\item  A fit should then be performed to minimize 
\begin{equation}
\chi ^{2}=\sum_{table-bins}\frac{\left( d\sigma _{\mu \mu -%
\mathop{\rm mod}%
}/dx~dy-d\sigma _{\mu \mu -data}^{+}/dx~dy\right) ^{2}}{\sigma
_{stat}^{2}+\sigma _{syst}^{2}}
\end{equation}
in each beam mode with respect to the desired parameters in $d\sigma _{c-%
\mathop{\rm mod}%
}/dx~dy~d\vec{p}_{c}.$

\item  The confidence level for the fit may be obtained by comparing the $%
\chi ^{2}$ to the sum of effective DOF for table bins used in the fit.
\end{itemize}

To provide further experimental information for model testing, the following
kinematic quantities are given along with the cross-section in each bin: \ $%
\left\langle E_{HAD}\right\rangle $, the mean visible hadronic energy; $%
\left\langle E_{\mu 2}\right\rangle $, the mean energy of the secondary
muon; $\left\langle p_{T2in}^{2}\right\rangle $, the mean square of the
secondary muon's transverse momentum in the event scattering plane; and$%
\left\langle p_{T2out}^{2}\right\rangle $, the mean square of the secondary
muons transverse momentum perpendicular to the event scattering plane. These
quantities are computed from the dimuon data, with the LO fit used only for
acceptance and smearing corrections. Tables~\ref{nutev nu bin1}-\ref{ccfr
nubar bin3} contain the measurements\footnote{%
These data may also be obtained electronically at www-e815.fnal.gov.}.
Appendix \ref{high-x-section} contains a supplementary discussion of the cross
sections at high $x$.

\section{Summary}

We present a measurement of the dimuon production cross-section from an
analysis of the data of the NuTeV neutrino DIS experiment at the Tevatron.
NuTeV data are combined with an earlier measurement from the CCFR experiment
that used the same detector but a different beamline. A leading order QCD
analysis of charm production performed on the combined data yields the
smallest errors to date on model parameters describing the charm mass, the
size and shape of the strange sea, and the mean semi-muonic branching
fraction of charm. The leading order QCD model describes NuTeV and CCFR data
very well, but cross-section model parameters extracted vary depending on
the particular choice of model. The extracted dimuon production
cross-section, by contrast, is insensitive to the choice of leading order
QCD model and should be of most use to the community of phenomenologists.

\appendix

\section{Cross-Section Table Binning Procedure}

\label{binstuff}

In this analysis we report the differential cross-sections $d\sigma \left(
\nu _{\mu }/\bar{\nu}_{\mu }Fe\rightarrow \mu ^{\mp }\mu ^{\pm }X\right)
/dxdy$ for forward secondary muons tabulated in bins of neutrino energy $E$,
Bjorken scaling variable $x$, and inelasticity $y$. The measurement is
obtained by using a LO Monte Carlo fit to the data to find the
correspondence (mapping) between the ``true'' (unsmeared) and reconstructed
(smeared) phase space, as discussed in Sec.~\ref{sec1}. This procedure maps
the statistical fluctuations of the observed event rate to the ``true''
phase space bins where the cross-section is reported. The consistency of the
procedure is checked by comparing fits of various models to the extracted
cross-section tables to fits of the same models performed directly to the
data. The criteria for the check is that the obtained central values and the
errors on the model parameters are the same in both cases. In order to meet
these criteria the binning of the smeared and the unsmeared phase space has
to be appropriately selected. 

Usually  in cross-section measurements, the procedure followed is to bin
both the ``true'' and the reconstructed variables using the same grid, and
select the bin size empirically so that for each bin the purity is maximized
and the smearing contribution from other bins is minimized\footnote{%
Purity is defined here as the fraction of events which have both their
unsmeared and smeared variables belonging to the same bin.}. In such a
method the correspondence between smeared and unsmeared bins is trivial
since the same grid of bins is used. In the case of the dimuon cross-section
measurement, a complication arises from the large smearing effects due to
the missing neutrino energy in the reconstructed dimuon final state. Unlike
detector resolution effects, this smearing is not a symmetric function of
the ``true,'' unsmeared variables, but rather an asymmetric mapping similar
to electroweak radiative corrections. Because of this effect, we have
followed a more elaborate procedure to map the visible phase space bins to
those of the ``true'' phase space. This procedure allows us to obtain the
desired high purity for each bin and also achieve stability of the result
independently of the binning choice. In addition to mapping, our procedure
allows us to take into account the significant bin-to-bin correlations which
arise from the large smearing corrections without having to construct a full
error matrix. This correlation matrix can be calculated, but it is too
unwieldy to be useful; and it is difficult to incorporate effects of
correlated systematic errors in a meaningful way. In our treatment, we
estimate an effective number of degrees of freedom which allows us to obtain
from an {\em uncorrelated} fit to the cross-section tables the same fit
parameter errors as the ones obtained by directly fitting to the data, using
the usual $\Delta \chi ^{2}=1$ definition.

We begin the description of the technique by defining more precisely the
factors $D_{i}$ and ${\cal N}_{i,\text{LO fit}}^{\prime }$ from Eq.~\ref{measured XC}.
In this equation, the expression on the left hand side represents the
measured cross-section for dimuon production as a function of $x$, $y$, and $%
E$ in bin $i$ of the true phase space (with the requirement $E_{\mu
2}>E_{\mu 2\min }$). On the right hand side, $D_{i}$ is the number of data
events in bin $i$ (which has to be determined using the Monte Carlo mapping
procedure), and ${\cal N}_{i,\text{LO fit}}^{\prime }$ is the number of events
predicted in bin $i$ by the LO fit. To estimate the number of data events
associated with bin $i$, we start by selecting bins in 
generated phase space $(x,y,E)_{i}$, requiring equal number of 
Monte Carlo events in each
projection of the true (generated) phase space, so the number of events in
each $(x,y,E)_{i}$ bin is approximately the same. The visible phase space is
divided using the same algorithm but with a much finer grid than the generated
space. The mapping matrix ${\cal M}_{ij}$ makes the correspondence between
visible and generated phase spaces: 
\[
{\cal M}_{ij}={\cal N}_{ij}/{\cal N}_{j},
\]
where ${\cal N}_{ij}$ is the number of Monte Carlo events generated in bin $i$
which end up in visible bin $j$ and ${\cal N}_{j}$ is the total number of Monte
Carlo events in visible bin j. The coverage fraction ${\cal C}$ in visible
space is defined as 
\begin{equation}
{\cal C}=\sum_{j}^{N({\cal C})}{\cal N}_{ij}/{\cal N}_{i},
\end{equation}
where ${\cal N}_{i}$ is the number of Monte Carlo events in generated bin $i$. In
the case of ${\cal C}=1$ the sum is performed over all visible bins;
otherwise, the summation goes over the bins with the highest ${\cal N}_{ij}/{\cal N}_{i}$
ratios until the desired fractional coverage is obtained. Using the above
definitions, and for a given coverage ${\cal C}$, we can define the number of data
events which belong to a given true phase space bin $i$ (where the
cross-section is reported) as 
\begin{equation}
D_{i}=\sum_{j\in {\cal C}}{\cal M}_{ij}\cdot D_{j}.
\end{equation}
The number of Monte Carlo events in this generated bin, for a given coverage
${\cal C}$, is redefined as 
\begin{equation}
{\cal N}_{i}^{\prime }=\sum_{j\in {\cal C}}{\cal M}_{ij}{\cal N}_{j}.
\end{equation}

The cross-section error for each bin $i$ should be proportional to the
visible events ``mapped'' in that bin, so the following expression is used
to assign it: 
\begin{equation}
\delta _{i}=\frac{d\sigma _{\mu \mu }(\nu _{\mu }N\rightarrow \mu ^{-}\mu
^{+}X)_{i}}{dx\ dy}/\sqrt{{\cal N}_{i}}.
\end{equation}
Note that the Monte Carlo statistics contribution comes from the total
number of Monte Carlo events generated in bin $i$. The multiplicative factor 
$D_{i}/{\cal N}_{i}^{\prime }$ in Eq.~\ref{measured XC} 
cancels out to first order
the model dependence of the extracted cross-section and approximately
transfers the statistical fluctuation in the visible phase space to the true
phase space.

\begin{figure}[tbph]
\centerline{
\psfig{figure=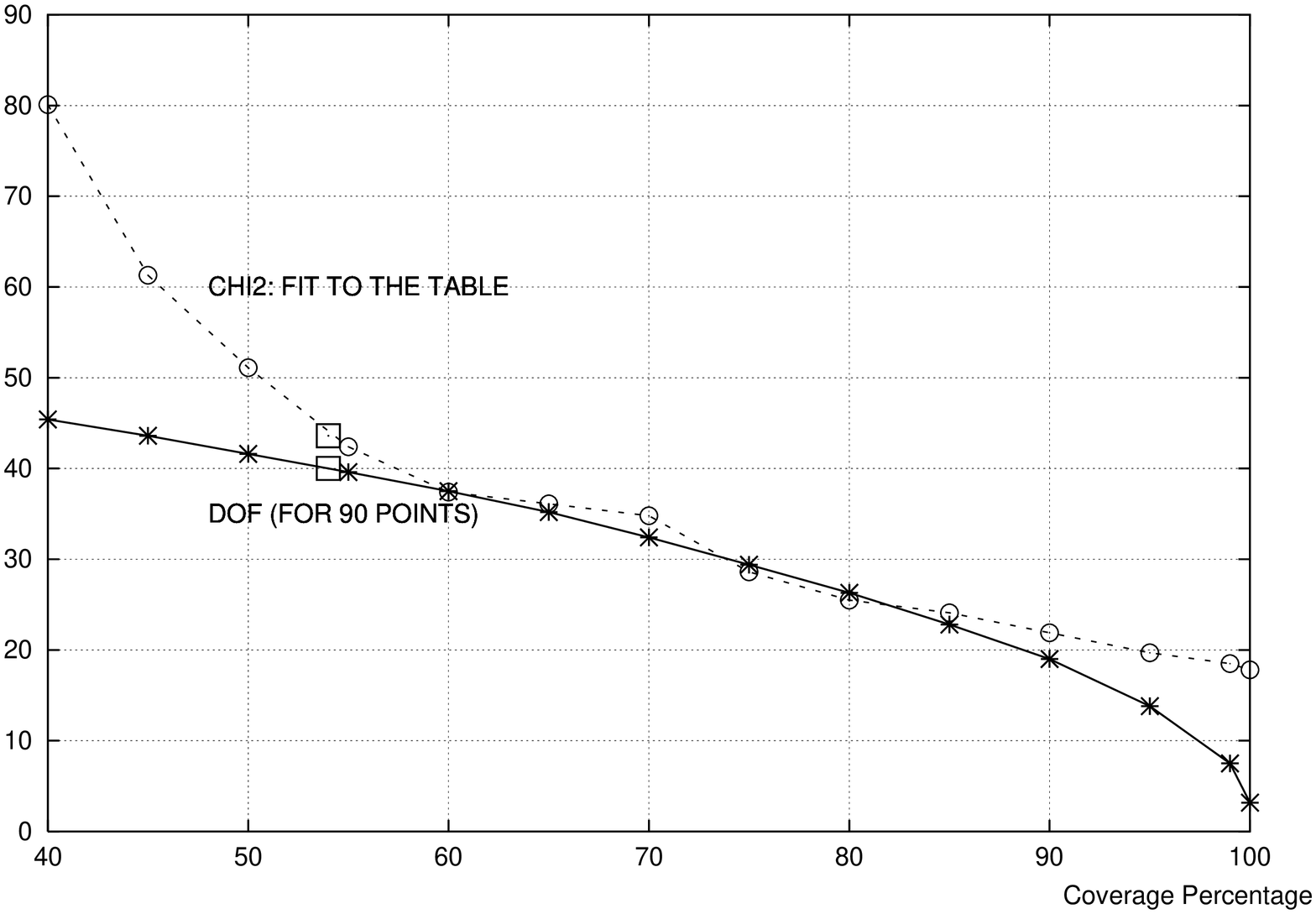,width=0.7\linewidth,height=0.7\linewidth,angle=-90}}
\caption{Effective DOF and $\protect\chi ^{2}$ as a function of coverage
area ${\cal C}$. The open symbols (dotted curve) show the $\protect\chi ^{2}$
obtained by fitting the cross-section table, while the stars (solid curve)
show the effective DOF. The open squares on both curves show the result
obtained by using the same grid on both smeared and unsmeared variables.}
\label{fig:dof}
\end{figure}

The procedure as described above is incomplete, since the true bins $i$ are
not statistically independent. As we stated in the introduction to this
section, we do not calculate the full error matrix, but rather estimate an
effective (independent) number of degrees of freedom. It is possible to
estimate the number of independent degrees of freedom by calculating the
contribution to the total number of degrees of freedom from each bin 
\begin{equation}
DOF_{i}=\frac{\sum_{i\in {\cal C}}{\cal M}_{ij}{\cal N}_{i}}{\sum_{j\in {\cal C}%
}{\cal N}_{j}}.
\end{equation}
It is obvious that the effective number of degrees of freedom depends on the
selected coverage fraction ${\cal C}$, and should decrease as ${\cal C}$
increases. Figure \ref{fig:dof} shows the number of effective degrees of
freedom (DOF) as a function of the coverage area ${\cal C}$ (solid curve).
The dotted curve in  Fig.~\ref{fig:dof} shows the $\chi ^{2}$ obtained
as a result of the fit to the table. We conclude that our method produces
the correct number for $\chi ^{2}/DOF$, if the effective number of degrees
of freedom is used, for a coverage fraction between $55\%$ and $90\%$.

The coverage area percentage used for our reported cross-section result is
based on a Monte Carlo study. In this study a Monte Carlo sample is used to
produce ``cross-section'' tables and then fits are performed to the tables
and directly to the sample. Using as guidelines the criteria that there
should be no pull on fit parameters in a fit to the cross-section table
versus a direct fit, and that an {\em uncorrelated} fit to the tables should
yield the same fit parameter errors as the ones obtained by a direct fit, we
selected a coverage area ${\cal C}=60\%$. The effective number of degrees of
freedom which corresponds to this value should be used with all fits
performed on the cross-section tables we present in this article.

\section{The high $x$ region}

\label{high-x-section}

This section presents a supplementary investigation of  
the high $x$ region ($x>0.5$).  The objective of this study is to 
ascertain whether there is any indication of an enhancement in the 
cross-section that we may be missing due to our use of wide $x$ bins
(dictated by the low observed event rate at high $x$).
Such an enhancement could be caused by an unusually large strange sea,
particularly in neutrino mode, which has been advocated to resolve
certain discrepancies between inclusive charged lepton and neutrino
scattering~\cite{barone2}.  Previous dimuon analyses~\cite{bazarko,rabinowitz}
may have missed  this effect due to dependence on the particular model
used to parameterize the strange sea distribution.

In order to minimize model-dependent corrections, we report our high x 
cross-section measurements as fractions of the total dimuon cross-section.  
Similarly, to quote a limit for the $x>0.5$ cross-section we use the 
observed data rate for $x_{VIS}>0.5$.  This is a conservative way to set a 
limit, since by the kinematic effect of the missing decay neutrino energy, 
the  contribution to a given $x_{VIS}$ bin always comes from $x<x_{VIS}$.

The cross-section ratio of the dimuon cross-section for $x>0.5$ to the total
dimuon cross-section in a given energy bin can be expressed as
\begin{equation}
\frac{\sigma^{2\mu}_{x>0.5}}{\sigma^{2\mu}} = \frac{N_{x>0.5}} {N_{tot}} \frac {M} {M_{E>5}},
\label{equ:highx}
\end{equation}
where $N_{x>0.5}$ is the number of observed events for $x_{VIS}>0.5$, 
$N_{tot}$ is the total number of observed dimuon events, $M$ is the Monte 
Carlo prediction with all experimental cuts applied, and $M_{E>5}$ is the Monte
Carlo prediction without the $\pi/K$ decay contribution and with only the 
$E_{\mu 2}>5$ GeV cut applied.  For this study, we use the same data selection
criteria described in section~\ref{dataselection}, exept for the  $x_{VIS}$
selection, which is changed to: $0\leq x_{VIS}\leq 2$.

In the anti-neutrino dimuon data sample we define three energy bins
and we record the number of the observed dimuon events with $x_{VIS}>0.5$ 
in the data, together with the Monte Carlo prediction and all the relevant
information for Eq.~\ref{equ:highx}.  
Here the Monte Carlo prediction is very well
constrained by our dimuon data in the full $x$ range, since the observed rate
is mostly due to scattering on $\bar{s}$ quarks. 
The results are presented in
Table~\ref{high-xb}.  
The systematic and statistical error 
on the  Monte Carlo prediction is negligible for this discussion.
Treating the Monte  Carlo prediction as a ``background,'' and using 
Eq.~\ref{equ:highx}, we set cross-section ratio upper limits at 90\% CL,
for any additional source of $x>0.5$ dimuons, 
of 0.0012, 0.007, and 0.009, respectively, in each one of the energy bins 
defined in Table~\ref{high-xb} (counting from lower to higher energy bin).

We follow the same procedure in the neutrino data sample.  Here, there is an
additional complication in the interpretation of the result since a large
contribution from valence quark  events is expected.  The Monte Carlo 
prediction for the rates of the non-strange sea component is not directly 
constrained by our dimuon data, but rather by 
inclusive structure function measurements.  
The results are presented in
Table~\ref{high-x}. 
It is worth noticing that
within the model and PDF sets used in this analysis (BGPAR) the Monte Carlo
prediction for the contribution of the strange sea is only on the order of 
3.7\% of the total rate for $x_{VIS}>0.5$; most of the contribution (69\%) 
in this model comes from the valence quarks. 

Using Eq.~\ref{equ:highx} and treating
the Monte Carlo prediction as a ``background'' source, we set 90\% CL limits 
for an additional cross-section source at $x>0.5$.  We find that for the first
and last energy bins in table~\ref{high-x} this additional source cannot be 
larger than 0.006 and
0.013 of the total dimuon cross-section, while for the 153.9-214.1 bin there
is less than 5\% probability that there is an additional source consistent with
our data (note that we have a 1.75$\sigma$ negative yield compared to our 
background prediction).

\begin{table}[tbp] \centering%
%
\begin{tabular}{|c|c|c|c|c|}
\hline
${\bf model}$ & ${\bf m}_{c}\left( \text{GeV}/c^{2}\right) $ & ${\bf %
\epsilon }$ & ${\bf B}_{c}\left( \%\right) $ & ${\bf \chi }^{2}/DOF$ \\ 
\hline
BGPAR & $1.33\pm 0.19\pm 0.10$ & $2.07\pm 0.31\pm 0.64$ & $11.40\pm 1.08\pm
1.15$ & $105/ 112$ \\ \hline
GRV & $1.65\pm 0.18\pm 0.09$ & $2.09\pm 0.31\pm 0.64$ & $11.11\pm 1.51\pm
1.60$ & $101/ 112$ \\ \hline
CTEQ & $1.63\pm 0.17\pm 0.09$ & $2.07\pm 0.31\pm 0.63$ & $10.70\pm 1.66\pm
1.76$ & $100/ 112$ \\ \hline
\end{tabular}

\begin{tabular}{|c|c|c|c|c|}
${\bf model}$ & ${\bf \kappa }$ & ${\bf \bar{\kappa}}$ & ${\bf \alpha }$ & $%
{\bf \bar{\alpha}}$ \\ \hline
BGPAR & $0.32\pm 0.06\pm 0.04$ & $0.37\pm 0.05\pm 0.04$ & $-1.10\pm 1.05\pm
0.59$ & $-2.78 \pm 0.42 \pm 0.40$ \\ \hline
GRV & $0.37\pm 0.05\pm 0.03$ & $0.37\pm 0.06\pm 0.06$ & $0.87\pm 1.25\pm
0.71 $ & $0.28 \pm 0.44 \pm 0.42$ \\ \hline
CTEQ & $0.44\pm 0.06\pm 0.04$ & $0.45\pm 0.08\pm 0.07$ & $1.17\pm 1.20\pm
0.68$ & $1.08 \pm 0.44 \pm 0.41$ \\ \hline
\end{tabular}
\caption{Results of LO fits to NuTeV data. The first error is statistical, the second systematic. 
\label{LO data fits}}%
\end{table}%
%

\begin{table}[tbp] \centering%
%
\begin{tabular}{|c|c|c|c|c|c|c|c|}
\hline
${\bf SET}$ & ${\bf m}_{c}\left( \text{GeV}/c^{2}\right) $ & ${\bf \epsilon }
$ & ${\bf B}_{c}\left( \%\right) $ & ${\bf \kappa }$ & ${\bf \bar{\kappa}}$
& ${\bf \alpha }$ & ${\bf \bar{\alpha}}$ \\ \hline
NuTeV & $1.33\pm 0.19$ & $2.07\pm 0.31$ & $11.40\pm 1.08$ & $0.32\pm 0.06$ & 
$0.37\pm 0.05$ & $-1.10\pm 1.05$ & $-2.78\pm 0.42$ \\ \hline
CCFR & $1.20\pm 0.23$ & $0.88\pm 0.12$ & $11.43\pm 0.95$ & $0.31\pm 0.05$ & $%
0.36\pm 0.05$ & $3.14\pm 0.91$ & $3.46\pm 0.73$ \\ \hline
combined & $1.38\pm 0.13$ & & $11.57\pm 0.70$ & $0.35\pm 0.04$
& $0.41\pm 0.04$ & $-0.77\pm 0.66$ & $-2.04\pm 0.36$ \\ \hline
\end{tabular}
\caption{Results of LO fits to NuTeV, CCFR, and combined data set, using the 
BGPAR PDF set.  Only the statistical errors are shown.
\label{set data fits}}%
\end{table}%
%

\begin{table}[tbp] \centering%
%
\begin{tabular}{|c|c|c|c|c|c|c|}
\hline
${\bf SET}$ & ${\bf m}_{c}\left( \text{GeV}/c^{2}\right) $ & ${\bf B}%
_{c}\left( \%\right) $ & ${\bf \kappa }$ & ${\bf \bar{\kappa}}$ & ${\bf %
\alpha }$ & ${\bf \bar{\alpha}}$ \\ \hline
NuTeV & $ 1.30\pm  0.22$ & $10.22\pm  1.11$ & $ 0.38\pm  0.08$ & $ 0.39\pm  0.06$ & $-2.07\pm  0.96$ & $-2.42 \pm  0.45 $ \\ \hline
CCFR  & $ 1.56\pm  0.24$ & $12.08\pm  0.99$ & $ 0.28\pm  0.05$ & $ 0.33\pm  0.05$ & $ 3.85\pm  1.17$ & $ 3.30 \pm  0.83 $ \\ \hline
NuTeV+& $ 1.40\pm  0.16$ & $11.00\pm  0.71$ & $ 0.36\pm  0.05$ & $ 0.38\pm  0.04$ & $-1.12\pm  0.73$ & $-2.07 \pm  0.39 $ \\ \hline
\end{tabular}
\caption{Results of LO fits to the cross-section tables extracted from the 
NuTeV, CCFR, and combined data sets.  
\label{set table fits}}%
\end{table}%
%

\begin{table}[tbp] \centering%
%
\begin{tabular}{|c|c|c|c|c|c|c|c|}
\hline
& ${\bf m}_{c}\left( \text{GeV}/c^{2}\right) $ & ${\bf \epsilon }$ & ${\bf B}%
_{c}\left( \%\right) $ & ${\bf \kappa }$ & ${\bf \bar{\kappa}}$ & ${\bf %
\alpha }$ & ${\bf \bar{\alpha}}$ \\ \hline
$\nu$ $\pi/K (15\%)$ & $0.022$ & $0.51$ & $0.81$ & $0.018$ & $0.031$ & $0.01$
& $0.05$ \\ \hline
$\bar \nu$ $\pi/K(21\%)$ & $0.006$ & $0.13$ & $0.06$ & $0.001$ & $0.017$ & $%
0.01$ & $0.17$ \\ \hline
$R_L (20\%)$ & $0.037$ & $0.09$ & $0.17$ & $0.001$ & $0.010$ & $0.48$ & $%
0.26 $ \\ \hline
$\mu$ energy scale $(1\%)$ & $0.080$ & $0.33$ & $0.74$ & $0.036$ & $0.023$ & 
$0.25$ & $0.24$ \\ \hline
Hadron energy scale $(0.4\%)$ & $0.012$ & $0.08$ & $0.02$ & $0.005$ & $0.003$
& $0.01$ & $0.04$ \\ \hline
MC statistics & $0.047$ & $0.02$ & $0.31$ & $0.012$ & $0.006$ & $0.23$ & $%
0.01$ \\ \hline
Flux & $0.010$ & $0.01$ & $0.07$ & $0.001$ & $0.000$ & $0.03$ & $0.03$ \\ 
\hline
Systematic Error & $0.104$ & $0.64$ & $1.15$ & $0.043$ & $0.043$ & $0.59$ & $%
0.40$ \\ \hline
\end{tabular}
\caption{Systematic error sources for the LO-QCD fit to the NuTeV 
data.\label{lo
systematics}}%
\end{table}%
%

\begin{table}[tbp] \centering%
%
\begin{tabular}{|c|c|c|c|c|c|c|c|}
\hline
${\bf \chi^2}$ & ${\bf x}$ & ${\bf y}$ & $\frac{d\sigma \left( \nu _{\mu
}N\rightarrow \mu ^{-}\mu ^{+}X\right) }{dxdy}|_{+}$ & $\left\langle
E_{HAD}\right\rangle $ & $\left\langle E_{\mu ^{+}}\right\rangle $ & $%
\left\langle p_{T\mu ^{+}in}^{2}\right\rangle $ & $\left\langle p_{T\mu
^{+}out}^{2}\right\rangle $ \\ \hline
$ 0.64$&$ 0.021$&$ 0.334$&$ 0.419\pm 0.071\pm 0.003$&$  18.3\pm  1.0$&$   7.9\pm  0.5$&$ 0.40\pm 0.55$&$ 0.14\pm  0.24$  \\ \hline
$ 0.45$&$ 0.058$&$ 0.334$&$ 0.538\pm 0.090\pm 0.022$&$  18.6\pm  1.2$&$   9.0\pm  0.6$&$ 1.08\pm 1.14$&$ 0.76\pm  1.49$  \\ \hline
$ 0.44$&$ 0.102$&$ 0.334$&$ 0.427\pm 0.069\pm 0.007$&$  17.9\pm  1.2$&$   8.3\pm  0.7$&$ 1.19\pm 0.78$&$ 0.26\pm  0.48$  \\ \hline
$ 0.46$&$ 0.168$&$ 0.334$&$ 0.323\pm 0.049\pm 0.008$&$  18.7\pm  1.2$&$   8.2\pm  0.5$&$ 0.74\pm 0.28$&$ 0.14\pm  0.25$  \\ \hline
$ 0.67$&$ 0.324$&$ 0.334$&$ 0.132\pm 0.019\pm 0.003$&$  19.4\pm  0.9$&$   8.5\pm  0.4$&$ 0.95\pm 0.09$&$ 0.07\pm  0.04$  \\ \hline
$ 0.64$&$ 0.021$&$ 0.573$&$ 0.774\pm 0.107\pm 0.015$&$  36.3\pm  1.4$&$  10.9\pm  0.8$&$ 0.19\pm 0.09$&$ 0.09\pm  0.03$  \\ \hline
$ 0.46$&$ 0.058$&$ 0.573$&$ 0.808\pm 0.108\pm 0.027$&$  34.3\pm  1.5$&$  10.9\pm  0.7$&$ 0.29\pm 0.14$&$ 0.11\pm  0.07$  \\ \hline
$ 0.47$&$ 0.102$&$ 0.573$&$ 0.792\pm 0.103\pm 0.012$&$  36.5\pm  1.5$&$  10.0\pm  0.6$&$ 1.08\pm 0.97$&$ 0.44\pm  0.96$  \\ \hline
$ 0.50$&$ 0.168$&$ 0.573$&$ 0.471\pm 0.060\pm 0.017$&$  34.8\pm  1.6$&$  10.6\pm  0.7$&$ 0.85\pm 0.72$&$ 0.25\pm  0.43$  \\ \hline
$ 0.62$&$ 0.324$&$ 0.573$&$ 0.198\pm 0.027\pm 0.003$&$  34.6\pm  1.6$&$  11.1\pm  0.7$&$ 0.74\pm 0.11$&$ 0.07\pm  0.03$  \\ \hline
$ 0.58$&$ 0.021$&$ 0.790$&$ 0.795\pm 0.126\pm 0.096$&$  49.6\pm  3.0$&$  14.6\pm  1.4$&$ 0.26\pm 0.35$&$ 0.10\pm  0.05$  \\ \hline
$ 0.43$&$ 0.058$&$ 0.790$&$ 0.894\pm 0.133\pm 0.029$&$  52.8\pm  2.5$&$  13.5\pm  1.5$&$ 0.34\pm 0.35$&$ 0.24\pm  0.44$  \\ \hline
$ 0.41$&$ 0.102$&$ 0.790$&$ 0.826\pm 0.123\pm 0.027$&$  52.1\pm  2.5$&$  11.2\pm  1.0$&$ 0.24\pm 0.21$&$ 0.09\pm  0.08$  \\ \hline
$ 0.52$&$ 0.168$&$ 0.790$&$ 0.706\pm 0.108\pm 0.005$&$  52.9\pm  2.5$&$  12.6\pm  1.2$&$ 0.30\pm 0.15$&$ 0.10\pm  0.08$  \\ \hline
$ 0.58$&$ 0.324$&$ 0.790$&$ 0.210\pm 0.038\pm 0.004$&$  49.0\pm  2.9$&$  12.8\pm  1.2$&$ 0.52\pm 0.18$&$ 0.09\pm  0.04$  \\ \hline
\end{tabular}
\caption{NuTeV forward differential cross-section for $\nu_\mu N \rightarrow \mu^- \mu^+ X$ at
$E\sim90.18$ GeV.  The forward cross-section requires $E_{\mu^+}\geq 5$ GeV, 
and
the cross-section values should be multiplied by $\frac{1}{100}\times G_{F}^{2}ME/\pi $. The first error
given for the cross-sections is statistical and the second systematic.
Units are in GeV or  GeV$^2$, where appropriate,  for the averages of the 
kinematic quantities.
\label{nutev nu bin1}}%
\end{table}%
%

\begin{table}[tbp] \centering%
%
\begin{tabular}{|c|c|c|c|c|c|c|c|}
\hline
${\bf \chi^2}$ & ${\bf x}$ & ${\bf y}$ & $\frac{d\sigma \left( \nu _{\mu
}N\rightarrow \mu ^{-}\mu ^{+}X\right) }{dxdy}|_{+}$ & $\left\langle
E_{HAD}\right\rangle $ & $\left\langle E_{\mu ^{+}}\right\rangle $ & $%
\left\langle p_{T\mu ^{+}in}^{2}\right\rangle $ & $\left\langle p_{T\mu
^{+}out}^{2}\right\rangle $ \\ \hline
$ 0.66$&$ 0.021$&$ 0.334$&$ 1.013\pm 0.148\pm 0.022$&$  39.1\pm  1.8$&$  12.1\pm  1.0$&$ 0.57\pm 1.32$&$ 0.24\pm  0.49$  \\ \hline
$ 0.45$&$ 0.058$&$ 0.334$&$ 0.837\pm 0.127\pm 0.018$&$  39.6\pm  2.1$&$  12.2\pm  1.1$&$ 0.33\pm 0.09$&$ 0.08\pm  0.04$  \\ \hline
$ 0.40$&$ 0.102$&$ 0.334$&$ 0.737\pm 0.111\pm 0.009$&$  39.8\pm  2.3$&$  12.2\pm  1.1$&$ 0.57\pm 0.17$&$ 0.08\pm  0.03$  \\ \hline
$ 0.39$&$ 0.168$&$ 0.334$&$ 0.484\pm 0.071\pm 0.010$&$  38.5\pm  2.2$&$  11.3\pm  0.9$&$ 0.69\pm 0.15$&$ 0.11\pm  0.10$  \\ \hline
$ 0.57$&$ 0.324$&$ 0.334$&$ 0.212\pm 0.028\pm 0.005$&$  39.5\pm  1.8$&$  12.7\pm  0.8$&$ 1.44\pm 0.20$&$ 0.08\pm  0.03$  \\ \hline
$ 0.57$&$ 0.021$&$ 0.573$&$ 1.304\pm 0.196\pm 0.015$&$  69.0\pm  2.3$&$  18.1\pm  1.8$&$ 0.19\pm 0.07$&$ 0.11\pm  0.06$  \\ \hline
$ 0.43$&$ 0.058$&$ 0.573$&$ 1.161\pm 0.176\pm 0.021$&$  71.0\pm  2.5$&$  16.7\pm  1.5$&$ 0.25\pm 0.08$&$ 0.08\pm  0.04$  \\ \hline
$ 0.35$&$ 0.102$&$ 0.573$&$ 1.140\pm 0.178\pm 0.019$&$  72.9\pm  3.2$&$  19.2\pm  2.1$&$ 0.63\pm 0.24$&$ 0.26\pm  0.78$  \\ \hline
$ 0.42$&$ 0.168$&$ 0.573$&$ 0.685\pm 0.107\pm 0.007$&$  75.0\pm  3.2$&$  14.8\pm  1.6$&$ 1.34\pm 1.26$&$ 0.15\pm  0.27$  \\ \hline
$ 0.52$&$ 0.324$&$ 0.573$&$ 0.242\pm 0.038\pm 0.004$&$  76.8\pm  3.3$&$  13.6\pm  1.6$&$ 0.74\pm 0.24$&$ 0.07\pm  0.04$  \\ \hline
$ 0.43$&$ 0.021$&$ 0.790$&$ 1.267\pm 0.179\pm 0.025$&$ 101.8\pm  3.6$&$  22.0\pm  2.8$&$ 0.20\pm 0.09$&$ 0.11\pm  0.05$  \\ \hline
$ 0.34$&$ 0.058$&$ 0.790$&$ 1.301\pm 0.176\pm 0.031$&$ 101.5\pm  3.2$&$  21.1\pm  2.4$&$ 0.29\pm 0.11$&$ 0.09\pm  0.04$  \\ \hline
$ 0.32$&$ 0.102$&$ 0.790$&$ 1.072\pm 0.153\pm 0.023$&$  98.6\pm  3.6$&$  21.9\pm  3.1$&$ 0.37\pm 0.17$&$ 0.09\pm  0.05$  \\ \hline
$ 0.37$&$ 0.168$&$ 0.790$&$ 0.788\pm 0.118\pm 0.020$&$ 101.8\pm  4.3$&$  20.4\pm  3.3$&$ 0.72\pm 0.65$&$ 0.09\pm  0.06$  \\ \hline
$ 0.41$&$ 0.324$&$ 0.790$&$ 0.251\pm 0.040\pm 0.005$&$ 100.4\pm  4.2$&$  20.2\pm  3.1$&$ 0.69\pm 0.40$&$ 0.08\pm  0.07$  \\ \hline
\end{tabular}
\caption{NuTeV forward differential cross-section for $\nu_\mu N \rightarrow 
\mu^- \mu^+ X$ at
$E\sim174.37$ GeV.  The forward cross-section requires $E_{\mu^+}\geq 5$ GeV, 
and
the cross-section values should be multiplied by $\frac{1}{100}\times G_{F}^{2}ME/\pi $. The first error
given for the cross-sections is statistical and the second systematic.
Units are in GeV or  GeV$^2$, where appropriate, for the averages of the
 kinematic
quantities.
\label{nutev nu bin2}}%
\end{table}%
%

\begin{table}[tbp] \centering%
%
\begin{tabular}{|c|c|c|c|c|c|c|c|}
\hline
${\bf \chi^2}$ & ${\bf x}$ & ${\bf y}$ & $\frac{d\sigma \left( \nu _{\mu
}N\rightarrow \mu ^{-}\mu ^{+}X\right) }{dxdy}|_{+}$ & $\left\langle
E_{HAD}\right\rangle $ & $\left\langle E_{\mu ^{+}}\right\rangle $ & $%
\left\langle p_{T\mu ^{+}in}^{2}\right\rangle $ & $\left\langle p_{T\mu
^{+}out}^{2}\right\rangle $ \\ \hline
$ 0.79$&$ 0.021$&$ 0.334$&$ 1.154\pm 0.167\pm 0.015$&$  54.8\pm  2.9$&$  15.1\pm  1.5$&$ 0.82\pm 1.29$&$ 0.09\pm  0.05$  \\ \hline
$ 0.54$&$ 0.058$&$ 0.334$&$ 1.306\pm 0.196\pm 0.034$&$  59.3\pm  3.1$&$  14.2\pm  1.3$&$ 0.31\pm 0.10$&$ 0.29\pm  0.40$  \\ \hline
$ 0.46$&$ 0.102$&$ 0.334$&$ 0.913\pm 0.132\pm 0.017$&$  61.2\pm  4.5$&$  13.2\pm  1.6$&$ 0.40\pm 0.11$&$ 0.11\pm  0.14$  \\ \hline
$ 0.41$&$ 0.168$&$ 0.334$&$ 0.712\pm 0.103\pm 0.013$&$  65.7\pm  3.3$&$  14.1\pm  1.4$&$ 0.82\pm 0.18$&$ 0.09\pm  0.03$  \\ \hline
$ 0.63$&$ 0.324$&$ 0.334$&$ 0.278\pm 0.036\pm 0.003$&$  69.9\pm  3.5$&$  13.8\pm  1.2$&$ 1.29\pm 0.30$&$ 0.11\pm  0.10$  \\ \hline
$ 0.68$&$ 0.021$&$ 0.573$&$ 1.487\pm 0.234\pm 0.010$&$  97.7\pm  3.8$&$  20.1\pm  3.0$&$ 0.19\pm 0.09$&$ 0.14\pm  0.16$  \\ \hline
$ 0.47$&$ 0.058$&$ 0.573$&$ 1.419\pm 0.235\pm 0.016$&$ 105.7\pm  3.9$&$  19.7\pm  2.6$&$ 0.83\pm 0.98$&$ 0.10\pm  0.09$  \\ \hline
$ 0.46$&$ 0.102$&$ 0.573$&$ 1.018\pm 0.166\pm 0.011$&$ 105.5\pm  6.1$&$  21.2\pm  3.9$&$ 0.84\pm 0.62$&$ 0.13\pm  0.52$  \\ \hline
$ 0.42$&$ 0.168$&$ 0.573$&$ 0.700\pm 0.116\pm 0.005$&$ 108.1\pm  4.9$&$  17.5\pm  3.4$&$ 0.78\pm 0.34$&$ 0.10\pm  0.07$  \\ \hline
$ 0.60$&$ 0.324$&$ 0.573$&$ 0.294\pm 0.046\pm 0.003$&$ 109.6\pm  4.5$&$  18.6\pm  2.7$&$ 0.86\pm 0.27$&$ 0.08\pm  0.04$  \\ \hline
$ 0.56$&$ 0.021$&$ 0.790$&$ 1.656\pm 0.222\pm 0.060$&$ 148.9\pm  4.9$&$  22.6\pm  3.3$&$ 0.19\pm 0.11$&$ 0.16\pm  0.20$  \\ \hline
$ 0.43$&$ 0.058$&$ 0.790$&$ 1.546\pm 0.216\pm 0.013$&$ 144.3\pm  5.5$&$  27.6\pm  4.5$&$ 0.39\pm 0.20$&$ 0.13\pm  0.17$  \\ \hline
$ 0.40$&$ 0.102$&$ 0.790$&$ 1.211\pm 0.174\pm 0.046$&$ 149.8\pm  7.2$&$  25.4\pm  4.4$&$ 1.25\pm 1.19$&$ 0.26\pm  0.51$  \\ \hline
$ 0.36$&$ 0.168$&$ 0.790$&$ 0.968\pm 0.141\pm 0.011$&$ 150.3\pm  6.9$&$  22.2\pm  3.5$&$ 0.52\pm 0.32$&$ 0.48\pm  1.98$  \\ \hline
$ 0.37$&$ 0.324$&$ 0.790$&$ 0.311\pm 0.047\pm 0.005$&$ 154.5\pm  6.3$&$  19.2\pm  3.6$&$ 0.50\pm 0.26$&$ 0.07\pm  0.04$  \\ \hline
\end{tabular}
\caption{NuTeV forward differential cross-section for $\nu_\mu N \rightarrow 
\mu^- \mu^+ X$ at
$E\sim244.72$ GeV.  The forward cross-section requires $E_{\mu^+}\geq 5$ GeV, 
and the 
cross-section values should be multiplied by $\frac{1}{100}\times G_{F}^{2}ME/\pi $. The first error
given for the cross-sections is statistical and the second systematic.
Units are in GeV or  GeV$^2$, where appropriate, for the averages of the
 kinematic
quantities.
\label{nutev nu bin3}}%
\end{table}%
%

\begin{table}[tbp] \centering%
%
\begin{tabular}{|c|c|c|c|c|c|c|c|}
\hline
${\bf \chi^2}$ & ${\bf x}$ & ${\bf y}$ & $\frac{d\sigma \left( \bar{\nu}_{\mu
}N\rightarrow \mu ^{+}\mu ^{-}X\right) }{dxdy}|_{+}$ & $\left\langle
E_{HAD}\right\rangle $ & $\left\langle E_{\mu ^{-}}\right\rangle $ & $%
\left\langle p_{T\mu ^{-}in}^{2}\right\rangle $ & $\left\langle p_{T\mu
^{-}out}^{2}\right\rangle $ \\ \hline
$ 0.61$&$ 0.016$&$ 0.356$&$ 0.403\pm 0.101\pm 0.004$&$  16.9\pm  1.3$&$   7.5\pm  0.8$&$ 0.11\pm 0.07$&$ 0.06\pm  0.06$  \\ \hline
$ 0.39$&$ 0.044$&$ 0.356$&$ 0.393\pm 0.101\pm 0.019$&$  15.9\pm  1.9$&$   8.2\pm  1.0$&$ 1.34\pm 1.80$&$ 0.59\pm  1.57$  \\ \hline
$ 0.34$&$ 0.075$&$ 0.356$&$ 0.417\pm 0.107\pm 0.007$&$  16.5\pm  1.6$&$   7.9\pm  1.1$&$ 0.42\pm 0.33$&$ 0.11\pm  0.13$  \\ \hline
$ 0.37$&$ 0.117$&$ 0.356$&$ 0.275\pm 0.067\pm 0.025$&$  16.5\pm  1.5$&$   8.4\pm  0.7$&$ 0.37\pm 0.19$&$ 0.12\pm  0.13$  \\ \hline
$ 0.63$&$ 0.211$&$ 0.356$&$ 0.080\pm 0.018\pm 0.003$&$  18.5\pm  1.9$&$   8.0\pm  0.9$&$ 0.74\pm 0.23$&$ 0.08\pm  0.06$  \\ \hline
$ 0.61$&$ 0.016$&$ 0.586$&$ 0.463\pm 0.097\pm 0.030$&$  30.6\pm  2.2$&$  13.0\pm  1.9$&$ 0.19\pm 0.09$&$ 0.05\pm  0.04$  \\ \hline
$ 0.39$&$ 0.044$&$ 0.586$&$ 0.533\pm 0.110\pm 0.007$&$  31.2\pm  2.5$&$  10.3\pm  1.4$&$ 0.17\pm 0.08$&$ 0.11\pm  0.05$  \\ \hline
$ 0.42$&$ 0.075$&$ 0.586$&$ 0.621\pm 0.125\pm 0.026$&$  33.5\pm  2.2$&$  10.7\pm  1.4$&$ 0.32\pm 0.17$&$ 0.05\pm  0.05$  \\ \hline
$ 0.44$&$ 0.117$&$ 0.586$&$ 0.357\pm 0.071\pm 0.008$&$  31.5\pm  2.4$&$  11.1\pm  1.3$&$ 0.35\pm 0.13$&$ 0.09\pm  0.06$  \\ \hline
$ 0.63$&$ 0.211$&$ 0.586$&$ 0.183\pm 0.036\pm 0.003$&$  32.2\pm  2.3$&$  10.7\pm  1.2$&$ 0.93\pm 0.39$&$ 0.07\pm  0.04$  \\ \hline
$ 0.53$&$ 0.016$&$ 0.788$&$ 0.623\pm 0.161\pm 0.009$&$  43.7\pm  3.8$&$  12.1\pm  1.6$&$ 0.14\pm 0.13$&$ 0.13\pm  0.20$  \\ \hline
$ 0.39$&$ 0.044$&$ 0.788$&$ 0.623\pm 0.154\pm 0.016$&$  44.8\pm  4.3$&$  10.1\pm  1.8$&$ 0.10\pm 0.08$&$ 0.04\pm  0.03$  \\ \hline
$ 0.42$&$ 0.075$&$ 0.788$&$ 0.520\pm 0.124\pm 0.004$&$  43.4\pm  3.7$&$  13.4\pm  3.3$&$ 0.19\pm 0.11$&$ 0.10\pm  0.09$  \\ \hline
$ 0.41$&$ 0.117$&$ 0.788$&$ 0.399\pm 0.098\pm 0.016$&$  41.1\pm  5.2$&$  11.9\pm  2.7$&$ 0.30\pm 0.21$&$ 0.09\pm  0.10$  \\ \hline
$ 0.64$&$ 0.211$&$ 0.788$&$ 0.153\pm 0.039\pm 0.004$&$  46.8\pm  5.1$&$  10.0\pm  2.4$&$ 0.30\pm 0.26$&$ 0.07\pm  0.06$  \\ \hline
\end{tabular}
\caption{NuTeV forward differential cross-section for $\bar{\nu}_\mu N 
\rightarrow \mu^+ \mu^- X$ at
$E\sim78.98$ GeV.  The forward cross-section requires $E_{\mu^-}\geq 5$ GeV, 
and the 
cross-section values should be multiplied by $G_{F}^{2}ME/\pi $. The first error
given for the cross-sections is statistical and the second systematic.
Units are in GeV or  GeV$^2$, where appropriate, for the averages of the
 kinematic
quantities.
\label{nutev nubar bin1}}%
\end{table}%
%

\begin{table}[tbp] \centering%
%
\begin{tabular}{|c|c|c|c|c|c|c|c|}
\hline
${\bf \chi^2}$ & ${\bf x}$ & ${\bf y}$ & $\frac{d\sigma \left( \bar{\nu}_{\mu
}N\rightarrow \mu ^{+}\mu ^{-}X\right) }{dxdy}|_{+}$ & $\left\langle
E_{HAD}\right\rangle $ & $\left\langle E_{\mu ^{-}}\right\rangle $ & $%
\left\langle p_{T\mu ^{-}in}^{2}\right\rangle $ & $\left\langle p_{T\mu
^{-}out}^{2}\right\rangle $ \\ \hline
$ 0.68$&$ 0.016$&$ 0.356$&$ 0.793\pm 0.171\pm 0.017$&$  32.8\pm  2.5$&$  12.7\pm  2.0$&$ 0.50\pm 0.62$&$ 0.78\pm  1.84$  \\ \hline
$ 0.42$&$ 0.044$&$ 0.356$&$ 0.744\pm 0.172\pm 0.024$&$  36.1\pm  3.3$&$  12.2\pm  2.0$&$ 0.23\pm 0.12$&$ 0.08\pm  0.06$  \\ \hline
$ 0.37$&$ 0.075$&$ 0.356$&$ 0.752\pm 0.174\pm 0.013$&$  33.6\pm  2.9$&$  11.2\pm  1.6$&$ 0.33\pm 0.11$&$ 0.21\pm  0.36$  \\ \hline
$ 0.38$&$ 0.117$&$ 0.356$&$ 0.427\pm 0.097\pm 0.009$&$  36.7\pm  3.7$&$  10.3\pm  1.4$&$ 0.73\pm 0.27$&$ 0.07\pm  0.06$  \\ \hline
$ 0.56$&$ 0.211$&$ 0.356$&$ 0.140\pm 0.030\pm 0.003$&$  38.5\pm  4.1$&$   8.8\pm  0.8$&$ 0.61\pm 0.22$&$ 0.06\pm  0.05$  \\ \hline
$ 0.53$&$ 0.016$&$ 0.586$&$ 1.338\pm 0.317\pm 0.025$&$  58.5\pm  3.0$&$  16.4\pm  2.2$&$ 0.09\pm 0.07$&$ 0.07\pm  0.04$  \\ \hline
$ 0.37$&$ 0.044$&$ 0.586$&$ 0.930\pm 0.223\pm 0.048$&$  66.3\pm  4.4$&$  13.2\pm  2.5$&$ 0.12\pm 0.09$&$ 0.09\pm  0.12$  \\ \hline
$ 0.36$&$ 0.075$&$ 0.586$&$ 0.744\pm 0.173\pm 0.017$&$  60.1\pm  4.5$&$  16.7\pm  2.5$&$ 0.27\pm 0.13$&$ 0.08\pm  0.08$  \\ \hline
$ 0.34$&$ 0.117$&$ 0.586$&$ 0.653\pm 0.159\pm 0.005$&$  64.0\pm  3.9$&$  14.7\pm  2.9$&$ 0.38\pm 0.18$&$ 0.08\pm  0.07$  \\ \hline
$ 0.50$&$ 0.211$&$ 0.586$&$ 0.257\pm 0.063\pm 0.006$&$  59.8\pm  4.6$&$  14.4\pm  3.2$&$ 0.69\pm 0.32$&$ 0.05\pm  0.04$  \\ \hline
$ 0.46$&$ 0.016$&$ 0.788$&$ 0.915\pm 0.198\pm 0.013$&$  83.0\pm  5.1$&$  18.1\pm  3.4$&$ 0.12\pm 0.09$&$ 0.09\pm  0.14$  \\ \hline
$ 0.37$&$ 0.044$&$ 0.788$&$ 1.106\pm 0.240\pm 0.042$&$  84.3\pm  6.4$&$  22.0\pm  4.1$&$ 0.30\pm 0.18$&$ 0.09\pm  0.07$  \\ \hline
$ 0.33$&$ 0.075$&$ 0.788$&$ 0.775\pm 0.172\pm 0.041$&$  87.3\pm  5.8$&$  18.4\pm  4.6$&$ 0.23\pm 0.31$&$ 0.07\pm  0.05$  \\ \hline
$ 0.36$&$ 0.117$&$ 0.788$&$ 0.547\pm 0.121\pm 0.015$&$  88.2\pm  6.5$&$  15.7\pm  4.6$&$ 0.17\pm 0.11$&$ 0.06\pm  0.06$  \\ \hline
$ 0.44$&$ 0.211$&$ 0.788$&$ 0.297\pm 0.075\pm 0.004$&$  78.7\pm  5.6$&$  17.3\pm  3.6$&$ 0.31\pm 0.13$&$ 0.07\pm  0.04$  \\ \hline
\end{tabular}
\caption{NuTeV forward differential cross-section for $\bar{\nu}_\mu N 
\rightarrow \mu^+ \mu^- X$ at
$E\sim146.06$ GeV.  The forward cross-section requires $E_{\mu^-}\geq 5$ GeV, 
and the 
cross-section values should be multiplied by $G_{F}^{2}ME/\pi $. The first error
given for the cross-sections is statistical and the second systematic.
Units are in GeV or  GeV$^2$, where appropriate, for the averages of the
 kinematic
quantities.
\label{nutev nubar bin2}}%
\end{table}%
%

\begin{table}[tbp] \centering%
%
\begin{tabular}{|c|c|c|c|c|c|c|c|}
\hline
${\bf \chi^2}$ & ${\bf x}$ & ${\bf y}$ & $\frac{d\sigma \left( \bar{\nu}_{\mu
}N\rightarrow \mu ^{+}\mu ^{-}X\right) }{dxdy}|_{+}$ & $\left\langle
E_{HAD}\right\rangle $ & $\left\langle E_{\mu ^{-}}\right\rangle $ & $%
\left\langle p_{T\mu ^{-}in}^{2}\right\rangle $ & $\left\langle p_{T\mu
^{-}out}^{2}\right\rangle $ \\ \hline
$ 0.80$&$ 0.016$&$ 0.356$&$ 1.046\pm 0.228\pm 0.015$&$  53.3\pm  3.3$&$  14.3\pm  2.4$&$ 0.16\pm 0.10$&$ 0.10\pm  0.05$  \\ \hline
$ 0.51$&$ 0.044$&$ 0.356$&$ 1.133\pm 0.254\pm 0.017$&$  55.0\pm  4.4$&$  15.9\pm  2.4$&$ 0.37\pm 0.23$&$ 0.08\pm  0.04$  \\ \hline
$ 0.41$&$ 0.075$&$ 0.356$&$ 0.855\pm 0.195\pm 0.010$&$  57.2\pm  6.4$&$  16.2\pm  3.5$&$ 0.52\pm 0.19$&$ 0.09\pm  0.07$  \\ \hline
$ 0.44$&$ 0.117$&$ 0.356$&$ 0.426\pm 0.094\pm 0.030$&$  57.1\pm  7.0$&$  13.6\pm  2.8$&$ 0.58\pm 0.37$&$ 0.07\pm  0.05$  \\ \hline
$ 0.60$&$ 0.211$&$ 0.356$&$ 0.331\pm 0.070\pm 0.029$&$  61.9\pm  6.9$&$  11.8\pm  2.6$&$ 1.46\pm 0.92$&$ 0.08\pm  0.04$  \\ \hline
$ 0.60$&$ 0.016$&$ 0.586$&$ 1.459\pm 0.380\pm 0.012$&$  95.0\pm  6.7$&$  17.2\pm  3.3$&$ 0.14\pm 0.11$&$ 0.04\pm  0.04$  \\ \hline
$ 0.45$&$ 0.044$&$ 0.586$&$ 1.111\pm 0.281\pm 0.011$&$  95.0\pm  5.4$&$  19.5\pm  3.9$&$ 0.27\pm 0.21$&$ 0.07\pm  0.05$  \\ \hline
$ 0.41$&$ 0.075$&$ 0.586$&$ 0.998\pm 0.253\pm 0.037$&$  89.2\pm  7.4$&$  25.3\pm  5.2$&$ 0.39\pm 0.22$&$ 0.15\pm  0.16$  \\ \hline
$ 0.42$&$ 0.117$&$ 0.586$&$ 0.787\pm 0.202\pm 0.011$&$  97.8\pm 10.3$&$  24.9\pm  7.9$&$ 1.24\pm 1.16$&$ 0.08\pm  0.10$  \\ \hline
$ 0.56$&$ 0.211$&$ 0.586$&$ 0.303\pm 0.077\pm 0.007$&$ 101.4\pm  6.3$&$  18.6\pm  5.8$&$ 1.17\pm 0.58$&$ 0.16\pm  0.33$  \\ \hline
$ 0.59$&$ 0.016$&$ 0.788$&$ 1.125\pm 0.243\pm 0.018$&$ 135.4\pm 12.9$&$  26.5\pm  6.1$&$ 0.15\pm 0.13$&$ 0.09\pm  0.09$  \\ \hline
$ 0.46$&$ 0.044$&$ 0.788$&$ 1.433\pm 0.295\pm 0.012$&$ 132.8\pm 10.2$&$  25.2\pm  6.0$&$ 0.17\pm 0.15$&$ 0.08\pm  0.11$  \\ \hline
$ 0.41$&$ 0.075$&$ 0.788$&$ 1.258\pm 0.268\pm 0.033$&$ 129.7\pm  9.7$&$  26.3\pm  7.3$&$ 0.52\pm 0.54$&$ 0.08\pm  0.09$  \\ \hline
$ 0.44$&$ 0.117$&$ 0.788$&$ 0.693\pm 0.154\pm 0.022$&$ 134.2\pm 14.4$&$  25.8\pm  7.4$&$ 0.70\pm 0.62$&$ 0.06\pm  0.07$  \\ \hline
$ 0.58$&$ 0.211$&$ 0.788$&$ 0.219\pm 0.050\pm 0.010$&$ 132.4\pm 18.0$&$  20.5\pm  5.3$&$ 0.68\pm 0.35$&$ 0.15\pm  0.17$  \\ \hline
\end{tabular}
\caption{NuTeV forward differential cross-section for $\bar{\nu}_\mu N 
\rightarrow \mu^+ \mu^- X$ at
$E\sim222.14$ GeV.  The forward cross-section requires $E_{\mu^-}\geq 5$ GeV, 
and the 
cross-section values should be multiplied by $G_{F}^{2}ME/\pi $. The first error
given for the cross-sections is statistical and the second systematic.
Units are in GeV or  GeV$^2$, where appropriate, for the averages of the
 kinematic
quantities.
\label{nutev nubar bin3}}%
\end{table}%
%

\begin{table}[tbp] \centering%
%
\begin{tabular}{|c|c|c|c|}
\hline
${\bf \chi^2}$ & ${\bf x}$ & ${\bf y}$ & $\frac{d\sigma \left( \nu _{\mu
}N\rightarrow \mu ^{-}\mu ^{+}X\right) }{dxdy}|_{+}$ \\ \hline
$ 0.54$&$ 0.023$&$ 0.320$&$ 0.589\pm 0.129\pm 0.018$  \\ \hline
$ 0.47$&$ 0.057$&$ 0.320$&$ 0.619\pm 0.110\pm 0.004$  \\ \hline
$ 0.44$&$ 0.100$&$ 0.320$&$ 0.472\pm 0.082\pm 0.004$  \\ \hline
$ 0.46$&$ 0.167$&$ 0.320$&$ 0.494\pm 0.081\pm 0.003$  \\ \hline
$ 0.69$&$ 0.336$&$ 0.320$&$ 0.159\pm 0.024\pm 0.001$  \\ \hline
$ 0.59$&$ 0.023$&$ 0.570$&$ 1.367\pm 0.223\pm 0.019$  \\ \hline
$ 0.43$&$ 0.057$&$ 0.570$&$ 1.017\pm 0.161\pm 0.008$  \\ \hline
$ 0.50$&$ 0.100$&$ 0.570$&$ 0.706\pm 0.103\pm 0.023$  \\ \hline
$ 0.52$&$ 0.167$&$ 0.570$&$ 0.472\pm 0.067\pm 0.010$  \\ \hline
$ 0.61$&$ 0.336$&$ 0.570$&$ 0.226\pm 0.034\pm 0.003$  \\ \hline
$ 0.42$&$ 0.023$&$ 0.795$&$ 1.406\pm 0.223\pm 0.047$  \\ \hline
$ 0.39$&$ 0.057$&$ 0.795$&$ 1.361\pm 0.203\pm 0.016$  \\ \hline
$ 0.46$&$ 0.100$&$ 0.795$&$ 1.138\pm 0.172\pm 0.014$  \\ \hline
$ 0.48$&$ 0.167$&$ 0.795$&$ 0.812\pm 0.136\pm 0.018$  \\ \hline
$ 0.49$&$ 0.336$&$ 0.795$&$ 0.211\pm 0.040\pm 0.003$  \\ \hline
\end{tabular}
\caption{CCFR E744/E770 forward differential cross-section for $\nu_\mu N 
\rightarrow \mu^- \mu^+ X$ at
$E\sim109.46$ GeV.  The forward cross-section requires $E_{\mu^+}\geq 5$ GeV, 
and the 
cross-section values should be multiplied by $G_{F}^{2}ME/\pi $. 
The first error
given for the cross-sections is statistical and the second systematic.
Units are in GeV or  GeV$^2$, where appropriate, for the averages of the
 kinematic
quantities.
\label{ccfr nu bin1}}%
\end{table}%
%

\begin{table}[tbp] \centering%
%
\begin{tabular}{|c|c|c|c|}
\hline
${\bf \chi^2}$ & ${\bf x}$ & ${\bf y}$ & $\frac{d\sigma \left( \nu _{\mu
}N\rightarrow \mu ^{-}\mu ^{+}X\right) }{dxdy}|_{+}$ \\ \hline
$ 0.58$&$ 0.023$&$ 0.320$&$ 1.319\pm 0.231\pm 0.008$  \\ \hline
$ 0.45$&$ 0.057$&$ 0.320$&$ 1.161\pm 0.192\pm 0.036$  \\ \hline
$ 0.46$&$ 0.100$&$ 0.320$&$ 0.932\pm 0.147\pm 0.005$  \\ \hline
$ 0.41$&$ 0.167$&$ 0.320$&$ 0.567\pm 0.090\pm 0.003$  \\ \hline
$ 0.68$&$ 0.336$&$ 0.320$&$ 0.254\pm 0.038\pm 0.004$  \\ \hline
$ 0.57$&$ 0.023$&$ 0.570$&$ 1.576\pm 0.248\pm 0.007$  \\ \hline
$ 0.43$&$ 0.057$&$ 0.570$&$ 1.709\pm 0.280\pm 0.004$  \\ \hline
$ 0.42$&$ 0.100$&$ 0.570$&$ 1.379\pm 0.226\pm 0.007$  \\ \hline
$ 0.43$&$ 0.167$&$ 0.570$&$ 0.916\pm 0.151\pm 0.004$  \\ \hline
$ 0.55$&$ 0.336$&$ 0.570$&$ 0.261\pm 0.043\pm 0.002$  \\ \hline
$ 0.45$&$ 0.023$&$ 0.795$&$ 1.642\pm 0.244\pm 0.047$  \\ \hline
$ 0.45$&$ 0.057$&$ 0.795$&$ 1.581\pm 0.241\pm 0.011$  \\ \hline
$ 0.38$&$ 0.100$&$ 0.795$&$ 1.092\pm 0.180\pm 0.004$  \\ \hline
$ 0.45$&$ 0.167$&$ 0.795$&$ 0.811\pm 0.134\pm 0.003$  \\ \hline
$ 0.54$&$ 0.336$&$ 0.795$&$ 0.228\pm 0.041\pm 0.003$  \\ \hline
\end{tabular}
\caption{CCFR E744/E770 forward differential cross-section for $\nu_\mu N 
\rightarrow \mu^- \mu^+ X$ at
$E\sim209.89$ GeV.  The forward cross-section requires $E_{\mu^+}\geq 5$ GeV, 
and the 
cross-section values should be multiplied by $G_{F}^{2}ME/\pi $. 
The first error
given for the cross-sections is statistical and the second systematic.
Units are in GeV or  GeV$^2$, where appropriate, for the averages of the
 kinematic
quantities.
\label{ccfr nu bin2}}%
\end{table}%
%

\begin{table}[tbp] \centering%
%
\begin{tabular}{|c|c|c|c|}
\hline
${\bf \chi^2}$ & ${\bf x}$ & ${\bf y}$ & $\frac{d\sigma \left( \nu _{\mu
}N\rightarrow \mu ^{-}\mu ^{+}X\right) }{dxdy}|_{+}$ \\ \hline
$ 0.85$&$ 0.023$&$ 0.320$&$ 1.546\pm 0.234\pm 0.005$  \\ \hline
$ 0.53$&$ 0.057$&$ 0.320$&$ 1.613\pm 0.256\pm 0.011$  \\ \hline
$ 0.53$&$ 0.100$&$ 0.320$&$ 1.014\pm 0.158\pm 0.017$  \\ \hline
$ 0.50$&$ 0.167$&$ 0.320$&$ 0.619\pm 0.095\pm 0.004$  \\ \hline
$ 0.71$&$ 0.336$&$ 0.320$&$ 0.293\pm 0.041\pm 0.003$  \\ \hline
$ 0.68$&$ 0.023$&$ 0.570$&$ 1.812\pm 0.300\pm 0.015$  \\ \hline
$ 0.50$&$ 0.057$&$ 0.570$&$ 1.690\pm 0.299\pm 0.006$  \\ \hline
$ 0.49$&$ 0.100$&$ 0.570$&$ 1.529\pm 0.271\pm 0.035$  \\ \hline
$ 0.50$&$ 0.167$&$ 0.570$&$ 0.756\pm 0.134\pm 0.003$  \\ \hline
$ 0.66$&$ 0.336$&$ 0.570$&$ 0.286\pm 0.048\pm 0.016$  \\ \hline
$ 0.72$&$ 0.023$&$ 0.795$&$ 2.422\pm 0.343\pm 0.027$  \\ \hline
$ 0.55$&$ 0.057$&$ 0.795$&$ 2.115\pm 0.331\pm 0.005$  \\ \hline
$ 0.55$&$ 0.100$&$ 0.795$&$ 1.689\pm 0.285\pm 0.005$  \\ \hline
$ 0.58$&$ 0.167$&$ 0.795$&$ 0.948\pm 0.159\pm 0.002$  \\ \hline
$ 0.76$&$ 0.336$&$ 0.795$&$ 0.328\pm 0.060\pm 0.002$  \\ \hline
\end{tabular}
\caption{CCFR E744/E770 forward differential cross-section for $\nu_\mu N 
\rightarrow \mu^- \mu^+ X$ at
$E\sim332.70$ GeV.  The forward cross-section requires $E_{\mu^+}\geq 5$ GeV, 
and the 
cross-section values should be multiplied by $G_{F}^{2}ME/\pi $. 
The first error
given for the cross-sections is statistical and the second systematic.
Units are in GeV or  GeV$^2$, where appropriate, for the averages of the
 kinematic
quantities.
\label{ccfr nu bin3}}%
\end{table}%
%

\begin{table}[tbp] \centering%
%
\begin{tabular}{|c|c|c|c|}
\hline
${\bf \chi^2}$ & ${\bf x}$ & ${\bf y}$ & $\frac{d\sigma \left( \bar{\nu}_{\mu
}N\rightarrow \mu ^{+}\mu ^{-}X\right) }{dxdy}|_{+}$ \\ \hline
$ 0.36$&$ 0.018$&$ 0.355$&$ 0.300\pm 0.138\pm 0.104$  \\ \hline
$ 0.41$&$ 0.041$&$ 0.355$&$ 0.470\pm 0.177\pm 0.006$  \\ \hline
$ 0.37$&$ 0.069$&$ 0.355$&$ 0.438\pm 0.154\pm 0.080$  \\ \hline
$ 0.41$&$ 0.111$&$ 0.355$&$ 0.259\pm 0.091\pm 0.088$  \\ \hline
$ 0.60$&$ 0.210$&$ 0.355$&$ 0.126\pm 0.042\pm 0.001$  \\ \hline
$ 0.24$&$ 0.018$&$ 0.596$&$ 1.135\pm 0.378\pm 0.057$  \\ \hline
$ 0.29$&$ 0.041$&$ 0.596$&$ 1.229\pm 0.393\pm 0.039$  \\ \hline
$ 0.31$&$ 0.069$&$ 0.596$&$ 0.569\pm 0.176\pm 0.132$  \\ \hline
$ 0.32$&$ 0.111$&$ 0.596$&$ 0.636\pm 0.205\pm 0.015$  \\ \hline
$ 0.50$&$ 0.210$&$ 0.596$&$ 0.181\pm 0.058\pm 0.009$  \\ \hline
$ 0.10$&$ 0.018$&$ 0.802$&$ 1.140\pm 0.404\pm 0.158$  \\ \hline
$ 0.14$&$ 0.041$&$ 0.802$&$ 0.875\pm 0.299\pm 0.086$  \\ \hline
$ 0.16$&$ 0.069$&$ 0.802$&$ 0.909\pm 0.320\pm 0.052$  \\ \hline
$ 0.30$&$ 0.111$&$ 0.802$&$ 0.872\pm 0.298\pm 0.036$  \\ \hline
$ 0.46$&$ 0.210$&$ 0.802$&$ 0.300\pm 0.115\pm 0.022$  \\ \hline
\end{tabular}
\caption{CCFR E744/E770 forward differential cross-section for 
$\bar{\nu}_\mu N \rightarrow \mu^+ \mu^- X$ at
$E\sim87.40$ GeV.  The forward cross-section requires $E_{\mu^-}\geq 5$ GeV, 
and the 
cross-section values should be multiplied by $G_{F}^{2}ME/\pi $. 
The first error
given for the cross-sections is statistical and the second systematic.
Units are in GeV or  GeV$^2$, where appropriate, for the averages of the
 kinematic
quantities.
%
\label{ccfr nubar bin1}}%
\end{table}%
%

\begin{table}[tbp] \centering%
%
\begin{tabular}{|c|c|c|c|}
\hline
${\bf \chi^2}$ & ${\bf x}$ & ${\bf y}$ & $\frac{d\sigma \left( \bar{\nu}_{\mu
}N\rightarrow \mu ^{+}\mu ^{-}X\right) }{dxdy}|_{+}$ \\ \hline
$ 0.32$&$ 0.018$&$ 0.355$&$ 0.944\pm 0.371\pm 0.043$  \\ \hline
$ 0.48$&$ 0.041$&$ 0.355$&$ 1.116\pm 0.386\pm 0.085$  \\ \hline
$ 0.37$&$ 0.069$&$ 0.355$&$ 0.854\pm 0.299\pm 0.018$  \\ \hline
$ 0.40$&$ 0.111$&$ 0.355$&$ 0.570\pm 0.191\pm 0.004$  \\ \hline
$ 0.57$&$ 0.210$&$ 0.355$&$ 0.234\pm 0.077\pm 0.017$  \\ \hline
$ 0.23$&$ 0.018$&$ 0.596$&$ 1.389\pm 0.489\pm 0.094$  \\ \hline
$ 0.37$&$ 0.041$&$ 0.596$&$ 1.542\pm 0.542\pm 0.038$  \\ \hline
$ 0.35$&$ 0.069$&$ 0.596$&$ 1.064\pm 0.374\pm 0.014$  \\ \hline
$ 0.35$&$ 0.111$&$ 0.596$&$ 0.718\pm 0.256\pm 0.010$  \\ \hline
$ 0.50$&$ 0.210$&$ 0.596$&$ 0.209\pm 0.071\pm 0.005$  \\ \hline
$ 0.11$&$ 0.018$&$ 0.802$&$ 1.699\pm 0.536\pm 0.243$  \\ \hline
$ 0.15$&$ 0.041$&$ 0.802$&$ 1.728\pm 0.592\pm 0.134$  \\ \hline
$ 0.27$&$ 0.069$&$ 0.802$&$ 1.734\pm 0.580\pm 0.065$  \\ \hline
$ 0.26$&$ 0.111$&$ 0.802$&$ 0.832\pm 0.287\pm 0.020$  \\ \hline
$ 0.32$&$ 0.210$&$ 0.802$&$ 0.217\pm 0.079\pm 0.009$  \\ \hline
\end{tabular}
\caption{CCFR E744/E770 forward differential cross-section for 
$\bar{\nu}_\mu N \rightarrow \mu^+ \mu^- X$ at
$E\sim160.52$ GeV.  The forward cross-section requires $E_{\mu^-}\geq 5$ GeV, 
and the 
cross-section values should be multiplied by $G_{F}^{2}ME/\pi $. 
The first error
given for the cross-sections is statistical and the second systematic.
Units are in GeV or  GeV$^2$, where appropriate, for the averages of the
 kinematic
quantities.
\label{ccfr nubar bin2}}%
\end{table}%
%

\begin{table}[tbp] \centering%
%
\begin{tabular}{|c|c|c|c|}
\hline
${\bf \chi^2}$ & ${\bf x}$ & ${\bf y}$ & $\frac{d\sigma \left( \bar{\nu}_{\mu
}N\rightarrow \mu ^{+}\mu ^{-}X\right) }{dxdy}|_{+}$ \\ \hline
$ 0.67$&$ 0.018$&$ 0.355$&$ 2.221\pm 0.714\pm 0.013$  \\ \hline
$ 0.54$&$ 0.041$&$ 0.355$&$ 1.165\pm 0.394\pm 0.015$  \\ \hline
$ 0.53$&$ 0.069$&$ 0.355$&$ 1.385\pm 0.451\pm 0.012$  \\ \hline
$ 0.48$&$ 0.111$&$ 0.355$&$ 0.583\pm 0.189\pm 0.005$  \\ \hline
$ 0.68$&$ 0.210$&$ 0.355$&$ 0.152\pm 0.049\pm 0.002$  \\ \hline
$ 0.45$&$ 0.018$&$ 0.596$&$ 2.774\pm 0.996\pm 0.120$  \\ \hline
$ 0.44$&$ 0.041$&$ 0.596$&$ 1.837\pm 0.673\pm 0.054$  \\ \hline
$ 0.43$&$ 0.069$&$ 0.596$&$ 1.319\pm 0.501\pm 0.020$  \\ \hline
$ 0.43$&$ 0.111$&$ 0.596$&$ 1.078\pm 0.415\pm 0.005$  \\ \hline
$ 0.58$&$ 0.210$&$ 0.596$&$ 0.234\pm 0.087\pm 0.006$  \\ \hline
$ 0.17$&$ 0.018$&$ 0.802$&$ 2.710\pm 0.796\pm 0.280$  \\ \hline
$ 0.28$&$ 0.041$&$ 0.802$&$ 1.567\pm 0.533\pm 0.061$  \\ \hline
$ 0.39$&$ 0.069$&$ 0.802$&$ 0.891\pm 0.334\pm 0.033$  \\ \hline
$ 0.46$&$ 0.111$&$ 0.802$&$ 0.643\pm 0.233\pm 0.006$  \\ \hline
$ 0.67$&$ 0.210$&$ 0.802$&$ 0.276\pm 0.103\pm 0.021$  \\ \hline
\end{tabular}
\caption{CCFR E744/E770 forward differential cross-section for 
$\bar{\nu}_\mu N \rightarrow \mu^+ \mu^- X$ at
$E\sim265.76$ GeV.  The forward cross-section requires $E_{\mu^-}\geq 5$ GeV, 
and the 
cross-section values should be multiplied by $G_{F}^{2}ME/\pi $. 
The first error
given for the cross-sections is statistical and the second systematic.
Units are in GeV or  GeV$^2$, where appropriate, for the averages of the
 kinematic
quantities.
\label{ccfr nubar bin3}}%
\end{table}%
%

\begin{table}[tbp] \centering%
%
\begin{tabular}{|c|c|c|c|c|c|c|}
${\bf E_{VIS}}$ & ${\bf N_{x>0.5}}$ & ${\bf MC}$ & ${\bf N_{\pi/k}}$ & $ {\bf N_{\bar{s}}}$ & $ {\bf N_{tot}}$ & ${\bf \frac{M}{M_{e>5}}}$ \\ \hline
34.8-128.6 & 1 & 3.4 & 2.1 & 1.3 & 688 & 0.67 \\ \hline
128.6-207.6 & 4 & 3.4 & 1.9 & 1.5 & 528 & 0.75 \\ \hline
207.6-388.0 & 2 & 3.4 & 2.1 & 1.3 & 238 & 0.78 \\ \hline
\end{tabular}
\caption{High-x events using the anti-neutrino
data sample.  $E_{VIS}$ is in GeV, 
$N_{x>0.5}$ is the number of observed events for $x_{VIS}>0.5$, $MC$ is the
Monte Carlo prediction for $x_{VIS}>0.5$, $N_{\pi/k}$ and $N_{\bar{s}}$ 
are the  $\pi/k$ decay and $\bar{s}$ contributions to $MC$, $N_{tot}$ is
the total number of dimuon events, and $\frac{M}{M_{e>5}}$ is the Monte Carlo
correction in Eq.~\ref{equ:highx}.
\label{high-xb}}%
\end{table}%
%

\begin{table}[tbp] \centering%
%
\begin{tabular}{|c|c|c|c|c|c|c|}
${\bf E_{VIS}}$ & ${\bf N_{x>0.5}}$ & ${\bf MC}$ & ${\bf N_{\pi/k}}$ & $ {\bf N_{s}}$ & $ {\bf N_{tot}}$ & ${\bf \frac{M}{M_{e>5}}}$ \\ \hline
36.1-153.9 & 65 & 53.39 & 11.6 & 2.0 & 2304 & 0.64 \\ \hline
153.9-214.1 & 42 & 53.44 & 14.8 & 1.8 & 1598 & 0.75 \\ \hline
214.1-399.5 & 60 & 53.38 & 17.4 & 2.2 & 1201 & 0.78 \\ \hline
\end{tabular}
\caption{High-x events using the neutrino
data sample.  $E_{VIS}$ is in GeV, 
$N_{x>0.5}$ is the number of observed events for $x_{VIS}>0.5$, $MC$ is the
Monte Carlo prediction for $x_{VIS}>0.5$, $N_{\pi/k}$ and $N_{s}$ 
are the  $\pi/k$ decay and $s$ contributions to $MC$, $N_{tot}$ is
the total number of dimuon events, and $\frac{M}{M_{e>5}}$ is the Monte Carlo
correction in Eq.~\ref{equ:highx}.
\label{high-x}}%
\end{table}%
%

\begin{figure}[tbp]
\centering{
\psfig{file=1muedst.epsi,width=1.\linewidth}
}
\caption{Charged-current Monte Carlo distributions (histograms) 
compared to data (points). 
Top: energy of the primary muon.
Middle: energy of the primary muon entering the toroid. Bottom: hadronic
energy. All energies are in GeV. Neutrino mode is shown on the left,
anti-neutrinos on the right side of the plot. The 
$\protect\chi^2$ per degree of freedom is
shown for each distribution.}
\label{fig:1mupl}
\end{figure}

\begin{figure}[tbp]
\centering{
\psfig{file=1muflux.epsi,width=1.\linewidth}
}
\caption{Charged-current Monte Carlo distributions (histograms)
compared to  data (points). Top: neutrino energy in GeV $E_{VIS} =
E_{1\protect\mu}+E_{2\protect\mu}+E_{had}$. Middle: event vertex position in
x plane. Bottom: event vertex position in y plane. Neutrinos are shown 
on the
left, anti-neutrinos on the right. The $\protect\chi^2$ per degree of
freedom is shown for each distribution.}
\label{fig:1muflux}
\end{figure}

\begin{figure}[tbp]
\centering{
\psfig{file=dimufit.epsi,width=1.\linewidth}
}
\caption{$x$, $z$, and $E$ distributions for dimuons (points) compared to
Monte Carlo (histogram). Neutrinos are shown
on the 
left, anti-neutrinos on the right. These distributions are used in the
log-likelihood fit.
The crosses show the Monte Carlo 
$\protect\pi K$ background component, and the stars the strange sea
contribution. }
\label{fig:dimufit}
\end{figure}

\begin{references}
\bibitem{rmp}  J.~M.~Conrad, M.~H.~Shaevitz, and T.~Bolton, Rev.~Mod.~Phys.\ {\bf 70}, 1341 (1998).

\bibitem{acot}  M.~A.~G.~Aivazis, F.~I.~Olness, and W.-K.~Tung, Phys.~Rev.~D 
{\bf 50}, 3085 (1994); M.~A.~G.~Aivazis, J.~C.~Collins, F.~I.~Olness, and W.-K.~Tung,
{\it ibid.}~{\bf 50}, 3102 (1994).

\bibitem{gkr}  M.~Gluck, S.~Kretzer, and E.~Reya, Phys.~Lett.~B {\bf 380},
171 (1997); {\it erratum ibid.} {\bf 405}, 391 (1997).

\bibitem{mrst}  A.~D.~Martin {\it et al.}, Eur.~Phys.~J.~C{\bf 2}, 287 (1998).

\bibitem{smith}  A.~Chuvakin, J.~Smith, and W.~L.~van~Neerven, Phys.~Rev.~D 
{\bf 61}, 096004 (2000).

\bibitem{barone}  V.~Barone {\it et al.}, Phys.~Lett.~B{\bf 328}, 143
(1994).

\bibitem{regina}  R.~Demina {\it et al.}, Phys.~Rev.~D {\bf 62}, 035011
(2000).

\bibitem{brodsky}  S.~J.~Brodsky and B.~Ma, Phys.~Lett.~B {\bf 381}, (317)
(1996).

\bibitem{pythia}  T.~Sjostrand, Comput.~Phys.~Commun. {\bf 82}, 74 (1994).

\bibitem{e531}  N.~Ushida, {\it et al.}, Phys.~Lett.~B{\bf 206}, 375 (1988); 
{\it ibid}, {\bf 206}, 380 (1988).

\bibitem{e531-tab}  T. Bolton, 1997, ``Determining the CKM Parameters V$%
_{cd} $ from $\nu $N Charm Production,'' KSU-HEP-97-04, e-print
hep-ex/9708014 (August, 1997).

\bibitem{PDG}  D.~E.~Groom {\it et al.}, Eur.~Phys.~Jour.~C {\bf 15}, 110
(2000).

\bibitem{bazarko}  A.~O.~Bazarko {\it et al.}, Z.~Phys.~C {\bf 65}, 189 (1995).

\bibitem{rabinowitz}  S.~A.~Rabinowitz {\it et al.},
Phys.~Rev.~Lett.~{\bf 70}, 134 (1993).

\bibitem{cdhs}  H.~Abramowicz {\it et al.}, Z.~Phys.~C {\bf 15}, 19 (1982).

\bibitem{charmii}  P.~Vilain {\it et al.}, Eur.~Phys.~J.~C {\bf 11}, 19 (1999).

\bibitem{nomad}  P.~Astier {\it et al.}, Phys.~Lett.~B {\bf 486}, 35 (2000).

\bibitem{collins}  P.~Collins and T.~Spiller, J.~Phys.~G {\bf 11}, 1289
(1985).

\bibitem{peterson}  C.~Peterson {\it et al.}, Phys.~Rev.~D{\bf 27}, 105
(1983).

\bibitem{drew}  A. Alton {\it et al.}, ``Search for
Light-to-Heavy Quark Flavor Changing Neutral Currents in $\nu _{\mu }N$ and $%
\bar{\nu}_{\mu }N$ Scattering at the Tevatron,'' KSU-HEP-00-001, 
e-print hep-ex/0007059, {\it to be
published in Phys.~Rev.~D} (July 31, 2000).

\bibitem{nim}  D.~A.~Harris {\it et al.}, Nucl.~Instrum.~Meth.~A {\bf 447}, 377 (2000).

\bibitem{geant}  R.~Brun {\it et al.}, ``GEANT Detector Description and
Simulation Library,'' CERN computing document CERN CN/ASD(1998).

\bibitem{bardin}  D.~Yu.~Bardin and O.~M.~Fedorenko, Sov.~J.~Nucl.~Phys. {\bf 30}, 418 (1979).

\bibitem{leplu}  G.~Ingelman, A.~Edin, and J.~Rathsman, Comput.~Phys.~Commun. {\bf 101}, 108 (1997).

\bibitem{pams}  P.~Sandler, ``Neutrino Production of Same-Sign Dimuons at
the Fermilab Tevatron,'' Ph.D. Thesis, Univ. of Wisconsin, 1992.

\bibitem{turtle}  D.~C.~Carey, F.~C.~Iselin, ``Decay TURTLE (Trace Unlimited Rays Through Lumped Elements): 
A Computer Program for Simulating Charged Particle Beam Transport
Systems, Including Decay Calculations,'' Fermilab preprint FERMILAB-PM-31 (1982).

\bibitem{atherton}  H.~W.~Atherton {\it et al.}, ``Precise Measurements Of
Particle Production by 400-GeV/c Protons in Beryllium Targets,'' CERN
preprint CERN-80-07 (1980).

\bibitem{malensek}  A.~J.~Malensek, ``Empirical Formula For Thick Target
Particle Production,'' Fermilab preprint FERMILAB-FN-0341 (1981).

\bibitem{grv94}  M.~Gluck, E.~Reya, and A.~Vogt, 
%
%
%
%
%
%
%
Z.~Phys.~C {\bf 67}, 433 (1995).

\bibitem{cteq5}  H.~L. Lai {\it et al.}, Eur.~Phys.~J.~C {\bf 12}, 375 (2000).

\bibitem{pdflib}  H.~Plothow-Besch, Comput.~Phys.~Commun.~{\bf 75}, 396
(1993).

\bibitem{bgpar}  A.~J.~Buras and K.~J.~F.~Gaemers, Nucl.~Phys.~B {\bf 132},
259 (1978).

\bibitem{whitlow}  L.~W.~Whitlow {\it et al.}, Phys.~Lett.~B {\bf 282}, 475
(1992).

\bibitem{barone2} V.~Barone, C.~Pascaud, and F.~Zomer, Eur.~Phys.~J.~C {\bf 12}, 243 (2000).
\end{references}
\end{document}